\pdfoutput=1 
\documentclass[preprint,5p,times,authoryear]{elsarticle}

\usepackage{amsmath}
\usepackage{caption}
\usepackage{array}
\usepackage{float}
\usepackage{mathtools}
\usepackage{tabularx}
\usepackage[usenames,dvipsnames]{xcolor}
\definecolor{forceblue}{rgb}{0.2235, 0.4157, 0.6941}
\usepackage[colorlinks]{hyperref}
\AtBeginDocument{%
	\hypersetup{
		citecolor=forceblue,
		linkcolor=forceblue,   
		urlcolor=forceblue}}
\usepackage{amsmath}
\usepackage{caption}
\usepackage{array}

\makeatletter
\def\ps@pprintTitle{%
	\let\@oddhead\@empty
	\let\@evenhead\@empty
	\def\@oddfoot{\footnotesize\itshape
		Preprint published in \ifx\@journal\@empty Elsevier
		\else\@journal\fi\hfill}
	\let\@evenfoot\@oddfoot
}
\makeatother

\journal{Journal of Wind Engineering and Industrial Aerodynamics 222 (2022) 104911 (\href{https://doi.org/10.1016/j.jweia.2022.104911}{doi.org/10.1016/j.jweia.2022.104911})}

\DeclareMathAlphabet{\mathcal}{OMS}{cmsy}{m}{n}
\SetMathAlphabet{\mathcal}{bold}{OMS}{cmsy}{b}{n}

\begin{document}
\begin{frontmatter}

\title{Data-driven Aerodynamic Analysis of Structures using Gaussian Processes}

\author[a]{Igor Kavrakov\corref{cor}}
\cortext[cor]{Corresponding author. Tel. +49 (0) 3643 584109}
\ead{igor.kavrakov@uni-weimar.de}
\author[b]{Allan McRobie}
\author[a]{Guido Morgenthal}
\address[a]{Chair of Modelling and Simulation of Structures, Bauhaus University Weimar, Marienstr. 13, Weimar 99423, Germany}
\address[b]{Department of Engineering, University of Cambridge, JJ Thomson Avenue 7a, Cambridge CB3 0FA, United Kingdom}

\begin{abstract}
An abundant amount of data gathered during wind tunnel testing and health monitoring of structures inspires the use of machine learning methods to replicate the wind forces. This paper presents a data-driven Gaussian Process-Nonlinear Finite Impulse Response (GP-NFIR) model of the nonlinear self-excited forces acting on structures. Constructed in a nondimensional form, the model takes the effective wind angle of attack as lagged exogenous input and outputs a probability distribution of the forces. The nonlinear input/output function is modeled by a GP regression. Consequently, the model is nonparametric, thereby circumventing to set up the function's structure a priori. The training input is designed as random harmonic motion consisting of vertical and rotational displacements. Once trained, the model can predict the aerodynamic forces for both prescribed input motion and aeroelastic analysis. The concept is first verified for a flat plate's analytical solution by predicting the self-excited forces and flutter velocity. Finally, the framework is applied to a streamlined and bluff bridge deck based on Computational Fluid Dynamics (CFD) data. The model's ability to predict nonlinear aerodynamic forces, flutter velocity, and post-flutter behavior are highlighted. Applications of the framework are foreseen in the structural analysis during the design and monitoring of slender line-like structures.

\end{abstract}
\begin{keyword}
Gaussian Processes \sep Data-driven \sep Structural Aerodynamics \sep Aeroelasticity \sep Machine Learning \sep Bridge Aerodynamics
\end{keyword}
\end{frontmatter}
\section{Introduction} \label{sec:Introduction}
This year marks the 8th decade since Tacoma Narrows Bridge collapsed due to wind, which arguably represents the starting point of modern structural aerodynamics. Since then, a significant progress has been made in this field, and one expects not to see another catastrophic failure as Tacoma's. However, bridge designers continually push the limits: They strive for longer and leaner structures such as the next longest span world record holder, the {\c C}anakkale 1915 Bridge. Having such structures also presents a need for more sophisticated aerodynamic models that capture nonlinear features as the wind forces are the leading action in the design. On another front, machine learning methods are becoming increasingly popular for reverse engineering. The application of these methods is based on one key ingredient that is abundant in wind engineering from wind tunnel tests and monitoring: data. This drives our motivation for of this paper, which purpose is to present a data-driven model of the wind forces acting on structures. \par
Traditionally, three types of wind models have been used to model the wind action on bluff bodies~\citep{Kavrakov4}: semi-analytical, numerical (based on Computational Fluid Dynamics (CFD)), and experimental. The semi-analytical models are white-box models based on analytical reasoning, and are supplemented by aerodynamic coefficients from a wind tunnel or CFD experiments to describe the aerodynamics. Many white-box type models have been developed through the years that can capture linear aerodynamics (cf. e.g. \cite{Davenport,Scanlan1978,Chen2000,Oiseth2011}), and nonlinear aerodynamics to a certain extent (cf. e.g.~\cite{Diana2013,WuKareem4,Skyvulstad2021}). While this type of model is commonly used for design purposes, they have limited predictive capabilities regarding nonlinear aerodynamics, such as higher-order harmonics, amplitude-dependence forces, and nonlinear fluid memory~\citep{KareemWu,AbbasKavrakovMorgenthal}. The CFD models utilize numerical methods to solve the Navier-Stokes equations, and have shown significant success in modeling the wind-structure interaction (cf. e.g., \cite{Sarkic2015,Nieto2015,Montoya2018,KavrakovMorgenthal3,MorgenthalGPU}). They still represent a major research direction in structural aerodynamics, particularly in three-dimensional applications. However, the relatively high model and numerical uncertainties of the CFD models hinder their reliability for practical bridge design due to high Reynolds numbers, and they are of high computational demand. Experimental models in wind tunnels are still used as the standard practice for designing long-span bridges, although they are prone to scaling issues and experimental uncertainty. Standard wind tunnel tests involve: i) sectional model tests to determine the aerodynamic coefficients including static wind coefficients, flutter derivatives and aerodynamic admittance functions; ii) full aeroelastic models. Typically, the static wind coefficients are determined for angles of attack up to $\pm$10 deg, while the flutter derivatives and aerodynamic admittance functions are determined about 0 deg angle of attack. For very long bridges, the aerodynamic admittance and flutter derivatives are also determined up to about $\pm$6 deg angle of attack, which are a necessary input for advanced nonlinear models such as, for e.g., the one by~\cite{Diana2020} based on the corrected quasi-steady theory. \par
The black-box data-driven models are receiving considerable attention in wind engineering. These models are attractive as the latent input/output function can be modeled using machine learning regression methods. This enables the data-driven models to capture various nonlinear aerodynamic features, typically intractable by the semi-analytical models. The computational cost of the data-driven models is relatively low compared to the CFD models. Moreover, such models can be trained not only on wind tunnel or CFD data but also on real monitoring data, leading to the potential to use them for real-time model updating~\citep{Castellon}. Several data-driven models have been recently presented to predict nonlinear aerodynamic forces, including flutter, and have been mainly based on Artificial Neural Networks~\citep{Wu2011,AbbasANN,LiWu}. These models include past terms of the deck displacements as input (typically in a dimensional form) and output the aerodynamic forces. Based on reasoning from system identification, models with such input (exogenous and lagged) can be classified as Nonlinear Finite Impulse Response (NFIR) models  ~\citep{Schoukens}. An NFIR model can capture a broad class of nonlinear systems with fading (i.e. finite) memory, which embodies the mathematical capabilities of some of the most complex semi-analytical aerodynamic models such as the Volterra's series~\citep{WuKareem4}.\par
\begin{figure}[!t]
	\centering
	\includegraphics[clip]{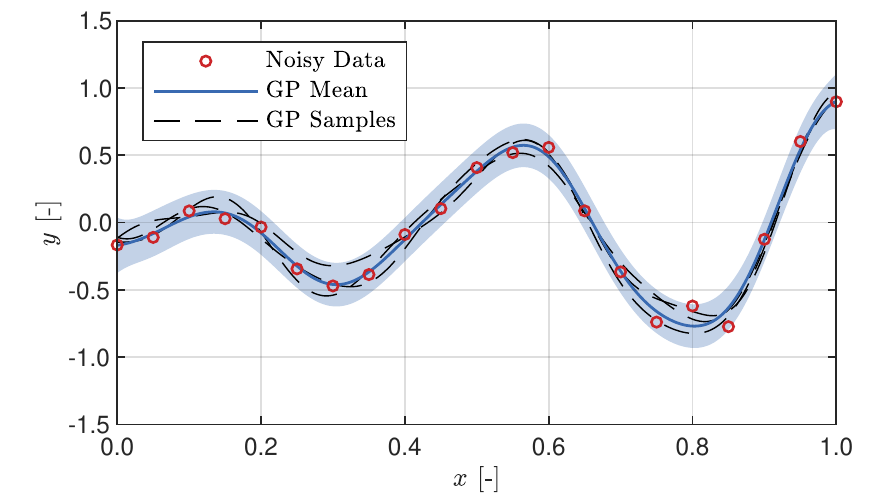} 
	\caption{GP prediction for noisy training data with samples drawn from the predictive distribution. The shaded area corresponds to the 99\% confidence interval.}
	\label{fig:GP_Plot}
\end{figure}

Another machine learning method that has found significant success for modeling dynamical systems is the Gaussian Processes (GPs) (cf. e.g.,~\cite{BookKocijan,Frigola2013,Deisenroth}). GPs can be viewed as a Bayesian nonparametric approach to regression, using a flexible distribution of functions as a prior. When conditioned on data, the GPs result in a robust posterior distribution that naturally quantifies the posterior uncertainty~(cf. Fig.~\ref{fig:GP_Plot}, \cite{BookRasmussen}). They can be viewed, with certain restrictions, as a prior on infinitely deep neural networks~\citep{Neal1996,lee2017deep}, with a powerful learning procedure that handles overfitting automatically. Since they are nonparametric, the model structure is not defined a priori, but it is governed by the data. For these reasons, they are particularly attractive to apply them in the field of structural aerodynamics.\par

This paper presents a novel reduced-order, data-driven GP-NFIR model for the nonlinear motion-induced forces acting on a bluff body such as a bridge deck. The input is formulated based on a a nondimensional form of the effective angle of attack, leading to an intuitive design of experiments for training and straightforward scaling between dimensions (e.g., wind tunnel model and real structure). An attractive nonparametric model form (in a statistical sense) is obtained by assuming that the latent function relating input to output is a GP, which bypasses the issues of designing a parametric structure in other data-driven aerodynamic models, such as e.g., the architecture for artificial neural networks. We first verify the model based on the linear analytical flat plate solution of the self-excited forces and critical flutter velocity. Finally, a practical application to bridge aerodynamics is presented in terms of predicting the nonlinear self-excited forces and post-flutter behavior, i.e. Limit Cycle Oscillations (LCO). The practical application data is generated using CFD; however, the presented framework is equally applicable for experimental data.\par

\section{GP-NFIR Aerodynamic Force Model}
\begin{figure}[!b]
	\centering
	\includegraphics[clip]{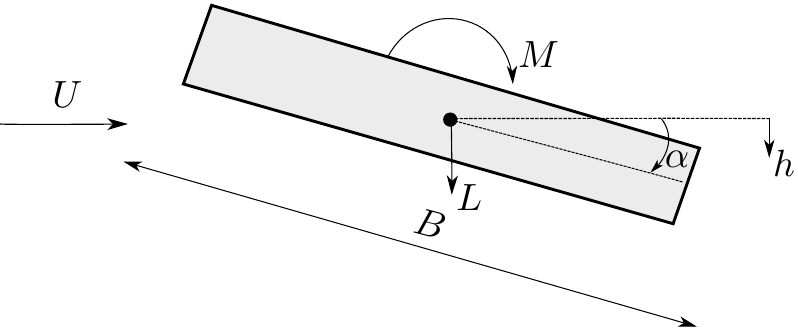} 
	\caption{Wind-structure interaction: Self-excited aerodynamic forces acting on an oscillating bluff body.}
	\label{fig:Forces}
\end{figure}
\subsection{Formulation}
Consider a two-dimensional (2D) wind-structure system (cf. Fig.~\ref{fig:Forces}) with two degrees of freedom (DOFs): vertical displacement $h=h(t)$ and torsional angle $\alpha=\alpha(t)$. The governing equations of motion for this system are:
\begin{equation}\label{eq:EqMot}
\begin{aligned}
m_h\ddot{h}+c_h\dot{h}+k_hh=L,\\
m_\alpha\ddot{\alpha}+c_\alpha\dot{\alpha}+k_\alpha \alpha=M,
\end{aligned}
\end{equation}
where $m_h$ and $m_\alpha$ is the mass, $c_h$ and $c_\alpha$ is the damping, and $k_h$ and $k_\alpha$ is the stiffness in the corresponding DOF.\par
The lift force $L=L(t)$ and moment $M=M(t)$ due to motion are given as:
\begin{equation}
\begin{aligned}
L=\frac{1}{2}\rho U^2BC_L(\alpha_e),\\
M=\frac{1}{2}\rho U^2B^2C_M(\alpha_e),
\end{aligned}
\end{equation}
where $\rho$ is the fluid density ($\rho$=1.2 $\mathrm{kg}/\mathrm{m}^3$ in this study), $U$ is the laminar free-stream velocity, $B$ is the chord, while $C_L$ and $C_M$ are the fluctuating lift and moment coefficients that are dependent on the effective angle of attack $\alpha_e=\alpha_e(t)$. In the absence of free-stream turbulence, the effective angle of attack is
\begin{equation}\label{eq:EffAngle}
\begin{aligned}
\alpha_e=\alpha+\arctan\left(\frac{\dot{h}+mB\dot{\alpha}}{U}\right),
\end{aligned}
\end{equation}
where $m$ describes the aerodynamic center.\par 
The semi-analytical models describe the lift and moment coefficients by taking the effective angle of attack (or its linearized version) as an input in a white-box manner (cf.e.g.~\cite{Scanlan1978,Chen2000,Diana2013,KavrakovMorgenthal,WuKareem4}). In turn, the data-driven black-box models describe the aerodynamic forces based on measured/simulated input/output samples. We formulate a data-driven aerodynamic force model as an NFIR model~\citep{Sjoberg} with additive independent identically distributed noise as:
\begin{equation}\label{eq:CoeffNFIR}
\begin{aligned}
C_{Li}&\displaystyle =f_L(\boldsymbol{\alpha}_i)+\epsilon_{Li},\\
C_{Mi}&=f_M(\boldsymbol{\alpha}_i)+\epsilon_{Mi},
\end{aligned}
\end{equation}

\noindent where $C_{Li}$ and $C_{Mi}$ are output samples (noisy observations) of the aerodynamic coefficients at step $i$, for $i=1,\dots,N_s$ and $N_s\in\mathbb{N}$ being the number of measured (training) samples, $f_L$ and $f_M$ are latent input/output functions, $\epsilon_{Li}\sim\mathcal{N}(0,\sigma_{nL}^2)$ and $\epsilon_{Mi}\sim\mathcal{N}(0,\sigma_{nM}^2)$ is the observation noise, and $\boldsymbol{\alpha}_i$ is the input sample vector. In an NFIR model, the input vector is constituted from current and past inputs, with which the system memory can be accounted for. We formulate the input vector as:
\begin{equation}\label{eq:Input}
\boldsymbol{\alpha}_i=(\alpha_{h,i}^\prime,\alpha_{a,i}^\prime,\alpha_{h,i},\alpha_{a,i},\alpha_{h,i-1},\alpha_{a,i-1},\dots{,\alpha_{h,i-S},\alpha_{a,i-S}}),
\end{equation}
for $S\in\mathbb{N}_0$ being the number of lag terms (past inputs). The components of the input vector are the angles related to the vertical displacements $\alpha_h=\arctan(\dot{h}/U)$ and rotation $\alpha_{a}=\alpha$ that constitute the effective angle of attack (cf.~\eqref{eq:EffAngle}). We describe these angles as functions of the nondimensional time $\alpha_h=\alpha_h(\tau)$ and $\alpha_a=\alpha_a(\tau)$ for $\tau=tU/B$. The derivative w.r.t. $\tau$ is denoted by a prime $(^\prime)$, and is related to the time derivative as: 
\begin{equation}
\begin{aligned}
\frac{\text{d}}{\text{d}\tau}=\frac{\text{dt}}{\text{d}\tau}\frac{\text{d}}{\text{d}t}=\frac{B}{U}\frac{\text{d}}{\text{d}t}.
\end{aligned}
\end{equation}
Based on this, the following relations can be obtained:
\begin{equation}
\begin{aligned}
\alpha_a^\prime&=\frac{B}{U}\dot{\alpha}, &&h^\prime=\frac{B}{U}{\dot{h}},\\
 \alpha_h&=\arctan\left(\frac{h^\prime}{B}\right),&&\alpha_h^\prime=\frac{1}{\displaystyle 1+\left(\frac{h^\prime}{B}\right)^2}\frac{h^{\prime\prime}}{B}.
\end{aligned}
\end{equation}\par
Generally, an NFIR model does not contain the input derivatives as they are implicitly included within the $i-1$ terms \citep{Schoukens}. In our formulation, we include the derivative of the effective angle only at step $i$ to give additional DOF since both the fluid memory and apparent mass effects are related to the velocity, as expressed in the analytical flat plate model~\citep{Theodorsen}. This also relates to the time-domain linear unsteady model in bridge aerodynamics (e.g.~\cite{Oiseth2011}), wherein the mathematical model of the impulse and indicial functions includes the derivative of the effective angle at step $i$ as a direct feed-through. Although, theoretically, adding the input derivatives should not make any difference, it showed in the application, empirically, that including this additional DOF in the data-driven model results in better numerical approximation.\par 
The next step is to formulate a statistical model for $f_L$ and $f_M$  (cf.~\eqref{eq:CoeffNFIR}) that can be trained from the available input/output data to predict the aerodynamic forces. To do so, we use GP regression, which edifice can be built on Bayesian nonparametric reasoning~\citep{BookRasmussen} where the priors of $f_L$ and $f_M$ are sets on an infinite-dimensional space of functions, and the posteriors are obtained by conditioning the priors on the available data. \par
The main assumption is that the latent functions $f_L$ and $f_M$ are independent GPs with zero means and covariance functions $k_L$ and $k_M$, i.e. $f_L\sim\mathcal{GP}(0,k_L)$ and $f_M\sim\mathcal{GP}(0,k_M)$. This assumption implies that the {\it prior} distributions of the function values $\boldsymbol{f}_L= f_L(\boldsymbol{\alpha})$ and $\boldsymbol{f}_M= f_M(\boldsymbol{\alpha})$ at finite input points $\boldsymbol{\alpha}\in\mathbb{R}^{N_s\times4+2S}$ are
\begin{equation}\label{eq:Prior}
\begin{aligned}
p(\boldsymbol{f}_L|\boldsymbol{\alpha})&\sim\mathcal{N}(\boldsymbol{0},\boldsymbol{K}_L),\\
p(\boldsymbol{f}_M|\boldsymbol{\alpha})&\sim\mathcal{N}(\boldsymbol{0},\boldsymbol{K}_M),
\end{aligned}
\end{equation}
where $\boldsymbol{K}_L= k_L(\boldsymbol{\alpha},\boldsymbol{\alpha})$ and $\boldsymbol{K}_M= k_M(\boldsymbol{\alpha},\boldsymbol{\alpha})$ are the covariance matrices that are obtained based on the corresponding covariance functions $k_L$ and $k_M$. A covariance function encodes the prior properties (i.e. assumptions) of the form of the latent function to be inferred (cf.~Sec.~\ref{sec:CovFunc} for details). The zero mean assumption in the prior~\eqref{eq:Prior} does not mean that the aerodynamic forces have zero mean, which is of course not true from a perspective of structural aerodynamics. Rather, it implies that the predictive mean will be captured by covariance functions in the posterior, as seen later (cf.~\eqref{eq:Prediction},~\eqref{eq:Pred}). \par
Having the aforementioned statistical assumptions in place, the {\it likelihood}, which is also Gaussian, can be written as
\begin{equation}\label{eq:likelihood}
\begin{aligned}
p(\boldsymbol{C}_L|\boldsymbol{f}_L,\boldsymbol{\alpha})&\sim\mathcal{N}(\boldsymbol{f}_L,\sigma_{nL}^2\boldsymbol{I}),\\
p(\boldsymbol{C}_M|\boldsymbol{f}_M,\boldsymbol{\alpha})&\sim\mathcal{N}(\boldsymbol{f}_M,\sigma_{nM}^2\boldsymbol{I}),
\end{aligned}
\end{equation}
where $\boldsymbol{C}_L$ and $\boldsymbol{C}_M$ are vectors containing the noisy observations (both are in $\mathbb{R}^{N_s\times 1}$), and $\boldsymbol{I}\in\mathbb{R}^{N_s\times N_s}$ is the identity matrix.\par
Using Bayes' theorem, the {\it posterior} is formulated as:
\begin{equation}\label{eq:Posterior}
\begin{aligned}
p(\boldsymbol{f}_L|\boldsymbol{C}_L,\boldsymbol{\alpha})&=\frac{p(\boldsymbol{C}_L|\boldsymbol{f}_L,\boldsymbol{\alpha})p(\boldsymbol{f}_L|\boldsymbol{\alpha})}{p(\boldsymbol{C}_L|\boldsymbol{\alpha})}\propto p(\boldsymbol{C}_L|\boldsymbol{f}_L,\boldsymbol{\alpha})p(\boldsymbol{f}_L|\boldsymbol{\alpha}),\\
p(\boldsymbol{f}_M|\boldsymbol{C}_M,\boldsymbol{\alpha})&=\frac{p(\boldsymbol{C}_M|\boldsymbol{f}_M,\boldsymbol{\alpha})p(\boldsymbol{f}_M|\boldsymbol{\alpha})}{p(\boldsymbol{C}_M|\boldsymbol{\alpha})}\propto p(\boldsymbol{C}_M|\boldsymbol{f}_M,\boldsymbol{\alpha})p(\boldsymbol{f}_M|\boldsymbol{\alpha}),\\
\end{aligned}
\end{equation}
which, considering only the proportional relation, can be analytically solved for the particular case of Gaussian likelihood (e.g. for the lift):
\begin{equation}\label{eq:Prediction}
p(\boldsymbol{f}_L|\boldsymbol{C}_L,\boldsymbol{\alpha})\sim\mathcal{N}(\boldsymbol{K}_L^T(\boldsymbol{K}_{L}+\sigma_{nL}^2\boldsymbol{I})^{-1}\boldsymbol{C}_L,\boldsymbol{K}_{L}-\boldsymbol{K}_{L}^T(\boldsymbol{K}_{L}+\sigma_{nL}^2\boldsymbol{I})^{-1}\boldsymbol{K}_{L}).\\ 
\end{equation} \par
We are generally interested in the prediction of the aerodynamic forces at angles of attack $\boldsymbol{\alpha}^*\in\mathbb{R}^{N_p\times 4+2S}$  (prediction samples) that are different from the training samples $\boldsymbol{\alpha}$. Using the same Baysian analogy like for the prediction at the training samples, the {\it predictive posteriors} of the aerodynamic forces for priors $p(\boldsymbol{f}_L|\boldsymbol{\alpha}^*)$ and $p(\boldsymbol{f}_L|\boldsymbol{\alpha}^*)$, and Gaussian likelihoods (cf.~\eqref{eq:likelihood}) can be obtained as:
\begin{equation}\label{eq:Pred}
\begin{aligned}
p(\boldsymbol{f}_L^*|\boldsymbol{\alpha}^*,\boldsymbol{C}_L,\boldsymbol{\alpha})&\propto p(\boldsymbol{C}_L|\boldsymbol{f}_L,\boldsymbol{\alpha})p(\boldsymbol{f}_L|\boldsymbol{\alpha}^*)\sim\mathcal{N}(\boldsymbol{m}_L^*,K_L^*),\\
p(\boldsymbol{f}_M^*|\boldsymbol{\alpha}^*,\boldsymbol{C}_M,\boldsymbol{\alpha})&\propto p(\boldsymbol{C}_M|\boldsymbol{f}_M,\boldsymbol{\alpha})p(\boldsymbol{f}_M|\boldsymbol{\alpha}^*)\sim\mathcal{N}(\boldsymbol{m}_M^*,K_M^*),\\
\end{aligned}
\end{equation}
with means $\boldsymbol{m}_L^*\in\mathbb{R}^{N_p\times 1}$ and $\boldsymbol{m}_M^*\in\mathbb{R}^{N_p\times 1}$:
\begin{equation}
\begin{aligned}\label{eq:PredMean}
\boldsymbol{m}_{L}^*&=\boldsymbol{K}_{LL^*}^T(\boldsymbol{K}_{L}+\sigma_{nL}^2\boldsymbol{I})^{-1}\boldsymbol{C}_L,\\
\boldsymbol{m}_{M}^*&=\boldsymbol{K}_{MM^*}^T(\boldsymbol{K}_{M}+\sigma_{nM}^2\boldsymbol{I})^{-1}\boldsymbol{C}_M,
\end{aligned}
\end{equation}
and covariances $\boldsymbol{K}_L^*\in\mathbb{R}^{N_p\times N_p}$ and $\boldsymbol{K}_M^*\in\mathbb{R}^{N_p\times N_p}$:
\begin{equation}\label{eq:PredVar}
\begin{aligned}
\boldsymbol{K}_{L}^*&=\boldsymbol{K}_{L^*}-\boldsymbol{K}_{LL^*}^T(\boldsymbol{K}_{L}+\sigma_{nL}^2\boldsymbol{I})^{-1}\boldsymbol{K}_{LL^*},\\
\boldsymbol{K}_{M}^*&=\boldsymbol{K}_{M^*}-\boldsymbol{K}_{MM^*}^T(\boldsymbol{K}_{M}+\sigma_{nM}^2\boldsymbol{I})^{-1}\boldsymbol{K}_{MM^*},
\end{aligned}
\end{equation}
where $\boldsymbol{K}_{L^*}= k_L(\boldsymbol{\alpha}^*,\boldsymbol{\alpha}^*)$, $\boldsymbol{K}_{M^*}= k_M(\boldsymbol{\alpha}^*,\boldsymbol{\alpha}^*)$, $\boldsymbol{K}_{LL^*}= k_L(\boldsymbol{\alpha},\boldsymbol{\alpha}^*)$ and $\boldsymbol{K}_{MM^*}= k_M(\boldsymbol{\alpha},\boldsymbol{\alpha}^*)$.\par
\begin{figure}[!t]
	\centering
	\includegraphics[clip]{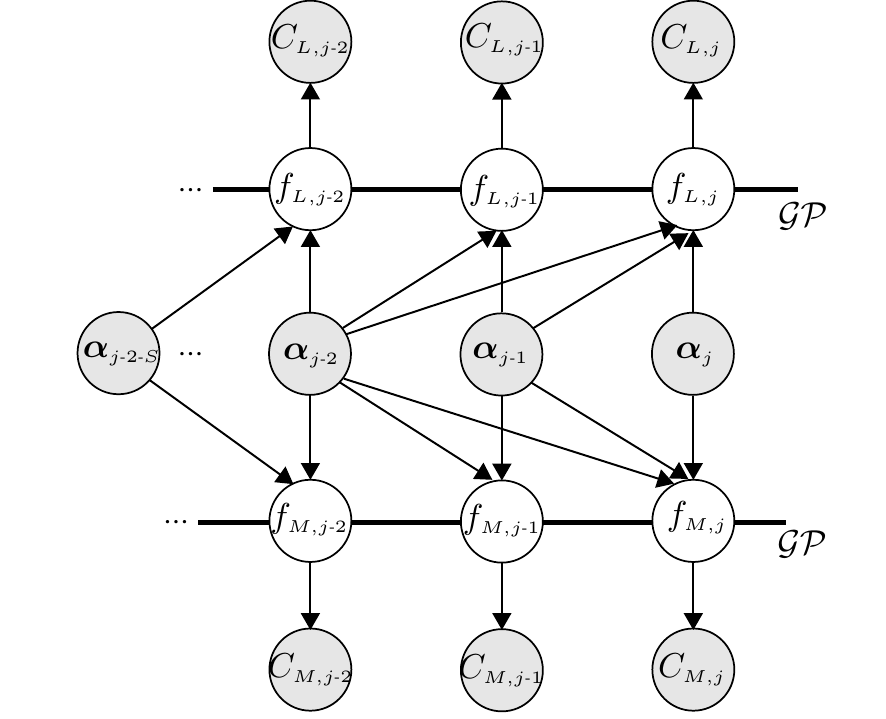} 
	\caption{Graphical representation of a GP-NFIR model of the aerodynamic forces (cf.~\eqref{eq:CoeffNFIR}): gray circles represent the observed variables and white circles represent the latent variables. The thick bar connects the latent variables that belong to the same GP.}
	\label{fig:ProbModel}
\end{figure}

This concludes the formulation of the aerodynamic forces as a data-driven GP-NFIR model~\citep{BookKocijan}. Figure~\ref{fig:ProbModel} depicts a graphical representation of this model with the relation between observed (in this figure, training) variables $\boldsymbol{\alpha}$, $\boldsymbol{C}_L$ and $\boldsymbol{C}_M$, and the latent variables $\boldsymbol{f}_L$ and $\boldsymbol{f}_M$. In case of a prediction at $\boldsymbol{\alpha}^*$, the joint distribution does not change, i.e. only the color of $\boldsymbol{C}_M^*$ and $\boldsymbol{C}_L^*$ circles becomes white. The model is completely nondimensional on the aerodynamic side, with only the effective angle as a relevant parameter in terms of amplitude and nondimensional frequency content (reduced frequency). This formulation makes for an easier transition of the learned latent functions $f_L$ and $f_M$ between scales (e.g., wind tunnel and real structure). If aeroelastic analyses are performed, the nondimensional time step should be selected considering the numerical algorithm for the dynamic time integration and the structural model. For comparison, other data-driven models (cf. e.g., ~\cite{Wu2011,AbbasANN}) are based on the nondimensional vertical amplitude and/or dimensional frequency. Further, these data-driven models can predict only the aerodynamic forces deterministically (i.e. the mean of the predictive distribution). The present GP-NFIR model additionally accounts for the noise in the force measurements (numerical or experimental) by assuming it is Gaussian while not assuming any probability distribution on the predictive mean w.r.t. the angle of attack. Currently, the lift and moment are considered as independent GPs to increase flexibility in the learning (e.g. the lift samples' quality does not influence the moment prediction) and reduce computational complexity in inverting the kernel. Further studies may construct a single model of lift and moment using multi-output GPs that will introduce further physical constrains in the modeling. The selection of covariance functions  $k_L$ and $k_M$, the learning of their hyperparameters, and the design of training input $\boldsymbol{\alpha}$ are discussed next. 
\subsection{Covariance function}\label{sec:CovFunc}
The covariance function (i.e. kernel) in the prior encodes the statistical model assumptions such as isotropy and invariances. Selecting an appropriate kernel is related to the problem at hand. Herein, we use the squared exponential covariance function with Automatic Relevance Determination (ARD), which relates the input sample vectors $\boldsymbol{\alpha}_i$ and $\boldsymbol{\alpha}_j$, both in $\mathbb{R}^{1\times4+2S}$, as
\begin{equation}\label{eq:Kernel}
k(\boldsymbol{\alpha}_i,\boldsymbol{\alpha}_j)=a^2\exp\left(-\frac{1}{2}(\boldsymbol{\alpha}_i-\boldsymbol{\alpha}_j)\boldsymbol{Q}(\boldsymbol{\alpha}_i-\boldsymbol{\alpha}_j){^T}\right),
\end{equation}
with variance $a^2$ and 
\begin{equation}\label{eq:Kernel2}
\boldsymbol{Q}=\mathrm{diag}(\boldsymbol{l}^{-2}),\hspace*{0.1cm}\text{for}\hspace*{0.1cm}\boldsymbol{l}=(l_{h^\prime},l_{a^\prime},l_{h,i},l_{a,i},\dots{l_{h,i-S},l_{a,i-S}}).
\end{equation}
The variance $a^2$ and the length scale vector $\boldsymbol{l}$ represent the kernel hyperparameters determined during learning. \par
The exponential kernel is smooth, stationary and a radial basis function. The ARD property~\citep{PhDNeal} accounts for the anisotropy in the input instead of having a single length scale for all input dimensions. This means that each length scale in the vector $\boldsymbol{l}$ separately scales the difference between the corresponding members of the input vectors $\boldsymbol{\alpha}_i$ and $\boldsymbol{\alpha}_j$ when computing the kernel. In other words, the vector $\boldsymbol{l}$ accounts for the sensitivity of the output w.r.t. each input dimension separately. Thus, it can be particularly useful for modeling aerodynamic forces due to the fluid memory effect, where the input at step $i$ is more relevant than at step e.g. $i-10$. Moreover, being infinitely differentiable, the exponential kernel has enjoyed particular success in GP modeling of dynamical systems~\citep{Frigola2013,Kocijan2005}, which is in line with the present application. The stationarity of the kernel for the present case is w.r.t. the input variable, i.e. the effective angle and its memory, which effectively warps the time variable~\citep{MacKay1998}. Thus, it can model potentially nonstationary aerodynamic forces and poses no symmetry constrain on the aerodynamic forces w.r.t. time.\par
The exponential kernel showed success for the present application; however, the question of which kernel is the most suitable for aerodynamic applications warrants further investigation. Alternatively, methods for automatic kernel constructions can be applied (cf. e.g., ~\cite{Calandra,PhDDuvenaud}).

\subsection{Learning}\label{sec:Learning}
Bayesian principles are again employed to learn the hyperparameters through maximizing the log {\it marginal likelihood} \citep{MacKay1992}. Example, given the observations of the lift force $\boldsymbol{C}_L$ for input $\boldsymbol{\alpha}$, we seek to minimize the log marginal likelihood:
\begin{equation}\label{eq:Minimisation}
\arg\max_{\boldsymbol{\theta}} \log p(\boldsymbol{C}_L|\boldsymbol{\alpha};\boldsymbol{\theta}),
\end{equation}
where the argument vector $\boldsymbol{\theta}=(a^2,\boldsymbol{l},\sigma_{nL}^2)$ contains the kernel hyperparameters, including the noise variance. For a GP model, the log marginal likelihood can be analytically derived \citep{BookRasmussen}:
\begin{equation}\label{eq:MarginalLikelihood}
\begin{aligned}
\log p(\boldsymbol{C}_L|\boldsymbol{\alpha};\boldsymbol{\theta})&=-\frac{1}{2}\boldsymbol{\alpha}^T(\boldsymbol{K}_{L}+\sigma_{nL}^2\boldsymbol{I})^{-1}\boldsymbol{\alpha}-\frac{1}{2}\log\left|\boldsymbol{K}_{L}+\sigma_{nL}^2\boldsymbol{I}\right|\\ &-\frac{N_p}{2}\log 2\pi.
\end{aligned}
\end{equation}\par
Typically, the training data for the aerodynamic forces contains a large number of samples. Such data can potentially lead to over-fitting and large errors for parametric data-driven models for effective angles of attack out of the training input space. In the case of GPs, obtaining the hyperparameters through the marginal likelihood automatically provides a trade-off between the data-fit and statistical model complexity. This trade-off is commonly referred to as Occam's razor, which helps prevent an overfit (cf.~\cite{RasmussenGhahramani} for an extensive discussion). \par

The marginal likelihood is a smooth function with multiple local minima. Although a global stochastic optimization would be more appropriate, we use a gradient-based solver that usually results in a local minimum. Using such solvers substantially decreases computational time, especially due to the possibility to obtain the analytical derivatives of the marginal likelihood w.r.t. the hyperparameters. The analytical partial derivatives of~\eqref{eq:MarginalLikelihood} are:
\begin{equation}\label{eq:LikelihoodDer}
\begin{aligned}
\frac{\partial}{\partial\theta_j}\log p(\boldsymbol{C}_L|\boldsymbol{\alpha};\boldsymbol{\theta})=\frac{1}{2}\mathrm{tr}&\Bigg(\boldsymbol{r}\boldsymbol{r}^T\frac{\partial(\boldsymbol{K}_{L}+\sigma_{nL}^2\boldsymbol{I}) }{\partial\theta_j}\\
&-(\boldsymbol{K}_{L}+\sigma_{nL}^2\boldsymbol{I})^{-1}\frac{\partial(\boldsymbol{K}_{L}+\sigma_{nL}^2\boldsymbol{I}) }{\partial\theta_j}\Bigg),
\end{aligned}
\end{equation}
where $\boldsymbol{r}=(\boldsymbol{K}_{L}+\sigma_{nL}^2\boldsymbol{I})^{-1}\boldsymbol{C}_L$ and $\mathrm{tr}(\cdot)$ is the trace. Even though only a local minimum is obtained, the resulting hyperparameters usually follow the ``good enough" principle after several reruns with different initial values.\par
For a large number of training samples, the computational complexity for inverting the covariance matrix in~\eqref{eq:MarginalLikelihood} is $\mathcal{O}(N_s^3)$; thus, the learning can be computationally demanding. To reduce the computational time only during learning, we use the method of a subset of data~\citep{BookRasmussen}. In this method, only a $N_s^*\in \mathbb{N}$ sample points are used, which are randomly selected from the training data set, i.e. $N_s^*\leq N_s$. The sample size can be selected through a factor $F=N_s/N_s^*$.

\begin{figure}[!t]
	\centering
	\includegraphics[clip]{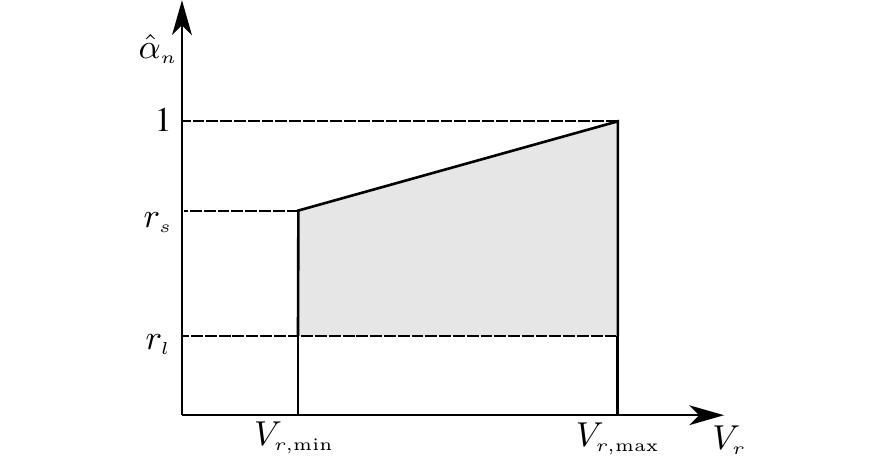} 
	\caption{Fourier amplitudes of a signal designed for training. The shaded area represents the sampling interval for the Fourier amplitudes $\hat{\alpha}_n$.}
	\label{fig:RandSig}
\end{figure}
\subsection{Design of experiments for training}\label{sec:DoE}
Data-driven models can be as good as their training data-set. A proper design of the effective angle of attack is warranted for training, such that the training data-set covers a wide range of frequencies and amplitudes. There are many types of excitation in nonlinear system identification, such as white noise, impulse, sinusoidal, quadratic chirp, or multi-harmonic excitation. From a practical aspect, the input signal should reflect the use of the data-driven model~\citep{Schoukens}. \par 
\begin{figure*}[!t]
	\centering
	\includegraphics[clip,width=\linewidth]{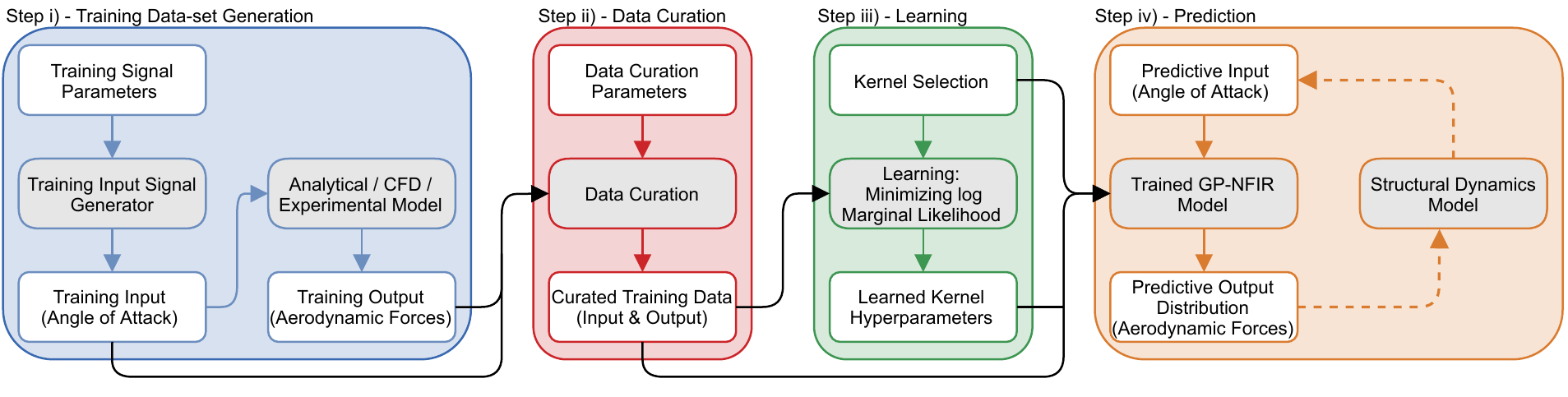} 
	\caption{Framework for prediction of the aerodynamic forces using a GP-NFIR model. A gray box indicates a model, while a white box indicates an input/output. The dashed lines in step iv) indicate if the aerodynamic forces are predicted step-by-step using the input (angle of attack) from a dynamic simulation (e.g. during flutter prediction). The structural dynamics model is obsolete in case of a multi-step ahead prediction using predefined angle of attack, i.e. forced-vibrations (other than the training). }
	\label{fig:Framework}
\end{figure*}
Since the motion (i.e. effective angle) in bridge aerodynamics is mainly harmonic during buffeting or flutter, we use a band-limited random harmonic signal with a specified standard deviation as training input. Taking $\alpha_h=\alpha_h(\tau)$ as an example, its normalized version $\alpha_{h,n}$ can be represented through random Fourier series as:
\begin{equation}\label{eq:InputTraining}
\alpha_{h,n}(\tau)=\int_{-\infty}^{\infty}\hat{\alpha}_{h,n}(V_r)\exp\left(i\frac{2\pi}{V_r}+i\varphi\right)\text{d}V_r,
\end{equation}
where $i$ is the imaginary unit, $V_r=U/(fB)$ is the reduced velocity, with $f$ being the oscillation frequency, and $\varphi\sim\mathcal{U}[0,2\pi)$ is a random, uniformly distributed phase for each harmonic. The random Fourier amplitude $\hat{\alpha}_{h,n}$ at each harmonic is obtained as:
\begin{equation}
\hat\alpha_{h,n}(V_r)\begin{cases}\sim\mathcal{U}[r_l,r_h],&\text{if}\ V_{r,\min}\leq V_r \leq V_{r,\max},\\=0,&\text{otherwise},\end{cases}
\end{equation}
where $V_{r,\min}$ and $V_{r,\max}$ define the selected range of reduced velocity for the training. The minimum relative Fourier amplitude at each $V_r$ is specified by the factor $0\leq r_l<1$, while the maximum by the factor
\begin{equation}
r_h(V_r)=r_{s}+\frac{V_r-V_{r,\min}}{V_{r,\max}-V_{r,\min}}(1-r_{s}),
\end{equation}
where $r_{s}$ defines the maximum relative Foruier amplitude at the lowest reduced velocity $V_r$, and can take a value between $r_{l}\leq r_{s}<1$. Figure~\ref{fig:RandSig} depicts schematically the region for sampling the random Fourier amplitude (gray area).\par
Finally, the training signal $\alpha_h$ is obtained by scaling:
\begin{equation}\label{eq:InputTrainingScaling}
\alpha_h=\alpha_{h,n}\frac{\sigma_{\alpha_{h}}}{\sigma_{\alpha_{h,n}}},
\end{equation}
where $\sigma_{\alpha_h}$ and $\sigma_{\alpha_{h,n}}$ are the standard deviations of the training (target) signal and its normalized version, respectively. 

The minimum/maximum relative Fourier amplitudes and reduced velocity range (cf. Fig.~\ref{fig:RandSig}, gray area) ensure that the sampled training signal contains the desired frequency content. By selecting a minimum relative Fourier amplitude (e.g. $r_l\approx 0.05$), it is ensured that all reduced velocities in the selected range ($V_{r,\min}$ - $V_{r,\max}$) are contained in the signal, i.e. no Fourier amplitude is zero. The maximum relative Fourier amplitude $r_s$ can be used to tune the high frequency components that may cause practical issues, i.e. forced excitation with large amplitudes for the high frequencies. Such excitation may cause numerical issues resulting in insufficiently resolved CFD simulations, or be mechanically infeasible in experiments. In the latter case, the signal should be tailored to be compatible with the arbitrary motion produced by an advanced forced-vibration rig (e.g.~\cite{Siedziako2017}), which is a specialized wind tunnel equipment. Typically, $r_s\approx0.2$ suffices for CFD simulations; however, this is dependent on minimum reduced velocity $V_{\min}$ and oscillation amplitude $\sigma_\alpha$. The reduced velocity range and the standard deviation of the amplitude should be carefully selected based on the expected oscillation range.\par

To cover both the linear and nonlinear range, two training signals with different amplitude range can be stacked together, yielding a new training signal   $\alpha_h=(\alpha_{h_1},\alpha_{h_2})$, where $\alpha_{h_1}$ and $\alpha_{h_2}$ have different standard deviations. When stacked together, it is important to note that the lagged terms at the interface between the two signals are taken as zero.

\begin{table*}[t]
	\centering
	\footnotesize
	\begin{tabularx}{\textwidth}{l | c  c |c c |c c c c c | c c | c c c c c}
		\hline
		
		\multicolumn{1}{c}{ }&\multicolumn{2}{c}{CFD Model}& \multicolumn{2}{c}{GP Model}& \multicolumn{5}{c}{Training Parameters} &\multicolumn{2}{c}{Flutter Der.} &\multicolumn{5}{c}{Dynamic Properties}  \\[2pt]
		Sec. &$\mathrm{Re}$ & $\Delta\tau_{\mathrm{CFD}}$  & $\Delta \tau_{\mathrm{GP}}$& $S$  & $V_{r,min} / V_{r,max}$ & $\tau$ & $\sigma_{\alpha_h}$ / $\sigma_{\alpha_a}$ & $r_l$ / $r_s$ & $F$ & $V_r$ & $\alpha_0$  & $m_h$ & $m_\alpha$ & $f_h$& $f_\alpha$&$\xi$ \\[2pt]
		&[-] & [-]  & [-] & [-]  & [-] & [-] & [deg] & [-] & [-] & [-] & [deg]& [t/m] & [tm$^2$/m]& [Hz] & [Hz] & [$\%$] \\[2pt]		
		
		\hline
		
		FP& / & /  & 0.05& 200  & 2 / 14 & 280 & 0.1 / 0.1 & 0.05 / 1.0 & 3 & 1-16 & 0.01  & 22.74 & 2470 & 0.1 & 0.278 & 0.3 \\[4pt]
		
		GB  & 1$\times$10$^5$ & 0.0165  & 0.05& 200  & 2 / 16 & 280 &  \parbox{1cm}{2.5 / 2.5, \\ 15 / 15} & 0.05 / 0.2 & 3 & 2-16 & 1  & 22.74 & 2470 & 0.1 & 0.278 & 0.5 \\ [6pt]
		
		H 	 & 5$\times$10$^4$ & 0.0083  & 0.05& 200 & 2 / 8 & 135 & \parbox{0.7cm}{1 / 1, \\ 8 / 8} & 0.05 / 0.2 & 3 & 2-7 & 3  & 4.25 & 177.73 & 0.13 & 0.2 & 1.0 \\[2pt]
		\hline
	\end{tabularx}
	\caption{Input parameters for training/prediction for the flat plate (FP), Great Belt's deck (GB) and H-shaped deck (H): $\mathrm{Re}$=Reynolds Number; $\Delta\tau_{\mathrm{CFD}}$=reduced CFD time step; $\Delta \tau_{\mathrm{GP}}$=reduced GP time step; $S$=number of lags; $V_{r,min},V_{r,max}$= minimum and maximum reduced velocities for training; $\tau$= training time; $\sigma_{\alpha_h}$,$\sigma_{\alpha_a}$=standard deviation of the angles for the training signal; $r_l,r_s$= relative Fourier amplitudes - minimum and maximum; $F$=subsampling factor; $V_r$=reduced velocities for prediction of flutter derivatives; $\alpha_0$ angle amplitude for prediction of flutter derivatives; $m_h$=vertical mass; $m_\alpha$=torsional mass; $f_h$=vertical frequency; $f_\alpha$=torsional frequency; $\xi$=damping ratio. For Great Belt's and H-shaped decks, two training signals with different standard deviations are stacked together.}
	\label{tab:NumParam}
\end{table*}

\begin{figure*}[!t]
	\centering
	\includegraphics[clip]{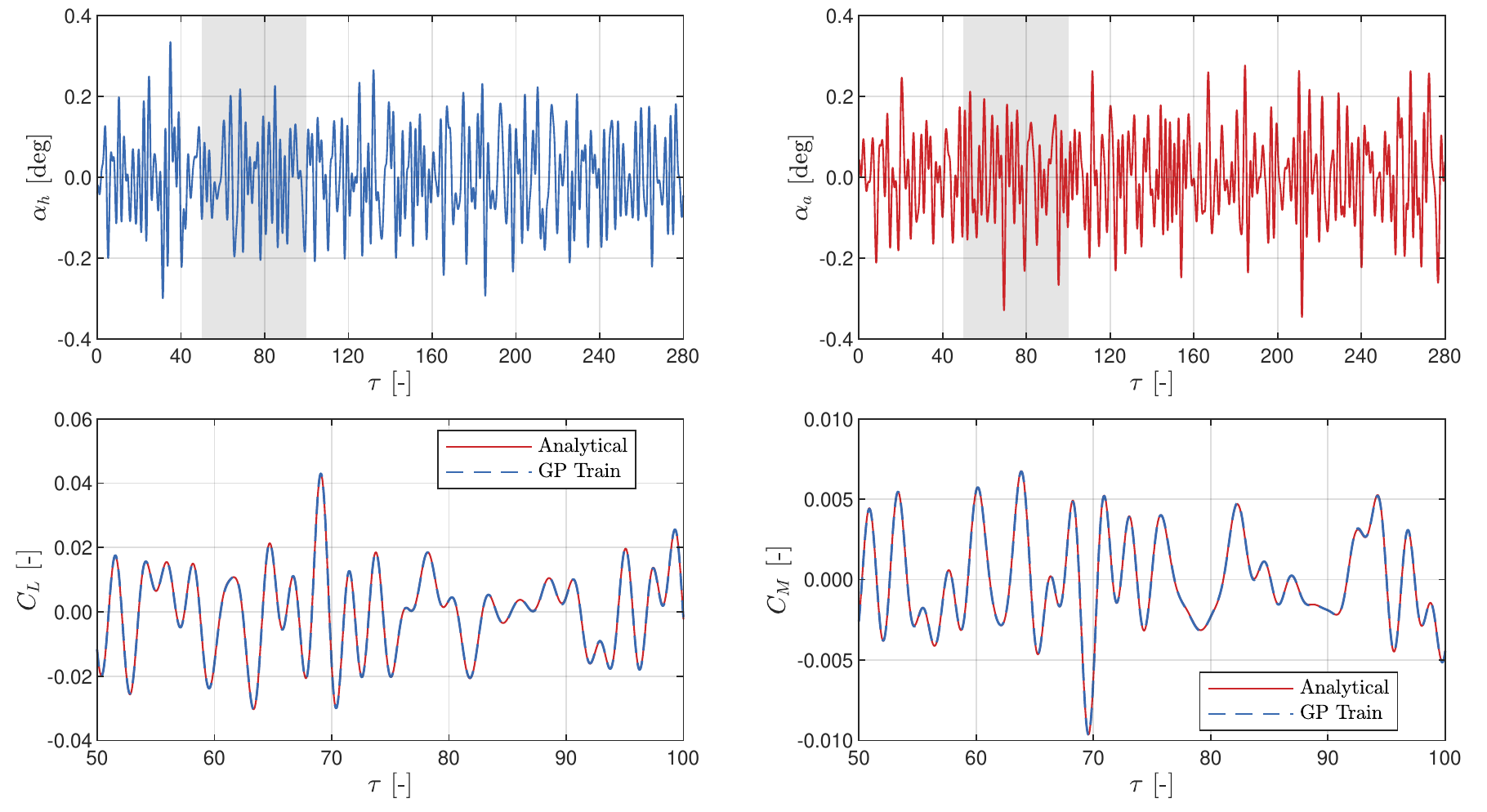} 
	\caption{Flat Plate: Training. Input random excitation for both DOFs (top) and output wind coefficients $C_L$ and $C_M$ (bottom). The mean of the GP posterior is represented by the dashed line, while the confidence interval is small (not visible). The wind coefficients (bottom) are depicted only for the corresponding shaded area of the input (top). The training range is for amplitudes with $\sigma_{\alpha_a}=\sigma_{\alpha_h}$=0.1 deg and reduced velocity range $V_r$=2-14.}
	\label{fig:Ex1a_Train}
\end{figure*}
\subsection{Framework, implementation and assessment}\label{sec:FrameWork}
The procedure to use the presented model for aeroelastic analyses consists of four straightforward steps (cf. Fig.~\ref{fig:Framework}):
  \begin{itemize}
 	\item[i)] Training data-set generation: select training signal parameters  ($r_l,r_s,V_{r,\min},V_{r,\max},\sigma_\alpha$), generate input training data $\boldsymbol{\alpha}$ (angle of attack, cf.~\eqref{eq:InputTraining}-\eqref{eq:InputTrainingScaling}), compute of output training data $\boldsymbol{C}_L$ and $\boldsymbol{C}_M$ (aerodynamic forces) based on forced-vibration simulation using an analytical/CFD/ experimental model;
 	\item[ii)] Data curation: Select data curation parameters (number of lags $S$, subsampling interval), curate training data (training data-set vector assembly (cf.~\eqref{eq:Input}), data normalization, subsampling (optional));
 	\item[iii)] Learning: select kernel $k$  (cf.~\eqref{eq:Kernel}-~\eqref{eq:Kernel2}), learn hyperparameters $\boldsymbol{\theta}$ (cf.~\eqref{eq:Minimisation}-\eqref{eq:LikelihoodDer}); 
 	\item[iv)] Prediction: predict the distribution of the aerodynamic forces ($p(\mathbf{f}_L^*)$ and $p(\mathbf{f}_M^*)$, cf.~\eqref{eq:Pred}-\eqref{eq:PredVar}) step-by-step ahead using predictive input from a dynamic simulation (cf.~\eqref{eq:EqMot}) or multi-step ahead using predefined predictive input (forced-vibration angle of attack $\boldsymbol{\alpha}^*$, other than the training $\boldsymbol{\alpha}$). 
 \end{itemize} \par 
This procedure is implemented in an independent Matlab code for the purpose of this study. The GP implementation is based on Algorithm 2.1 by~\cite{BookRasmussen}, which involves a Cholesky decomposition for the inversion of the covariance matrix in steps iii) and iv). Matlab's \verb|fminunc| function is used for the gradient-based optimization in step iii), including user-supplied derivatives of the marginal likelihood. The code for the flat plate example (cf. Sec.~\ref{sec:FundApp}) is included in an open-source repository: \href{https://github.com/IgorKavrakov/AeroGP}{github.com/IgorKavrakov/AeroGP}.\par
To assess the prediction of the aerodynamic forces of the GP-NFIR model, we use a set of nine comparison metrics for time-histories that are presented in \cite{KavrakovKareem} (including an open-source Matlab code). Here, we briefly list and revisit these metrics. Unless noted otherwise, during the analyses, we use similar metric parameters as in the study above, where further information can be found. \par 
For two signals $x=x(t)$ and $y=y(t)$, for $x$ being the reference signal, a comparison metric $\mathcal{M}^{x,y}=\mathcal{M}(x,y)$ is constructed as follows:
\begin{equation}
\mathcal{M}^{x,y}=\exp(-\lambda \mathcal{A}),
\end{equation}
where $\mathcal{A}=\mathcal{A}(x,y)$ is the relative exponent that quantifies a relative discrepancy in a particular signal feature, and $\lambda$ is the sensitivity parameter (we take $\lambda=1$ for all metrics in this study). With this, a comparison metric takes a value between 0$\leq\mathcal{M}\leq$1. A metric $\mathcal{M}=$1 indicates a complete match of the studied signal feature. This set of comparison metrics include a phase $\mathcal{M}_{\phi}$, a peak $\mathcal{M}_p$, a root-mean-square (RMS) $\mathcal{M}_{\mathrm{rms}}$, a magnitude $\mathcal{M}_{\mathrm{m}}$, a probability density function (PDF) $\mathcal{M}_{\mathrm{pdf}}$, a wavelet $\mathcal{M}_{w}$, a frequency-normalized wavelet $\mathcal{M}_{wf}$, a stationarity $\mathcal{M}_{s}$, and a bispectrum $\mathcal{M}_{b}$ metric. The phase metric $\mathcal{M}_{\phi}$ quantifies the mean phase of the signals through their cross-correlation, based on a relative phase normalization time $T_c$ that is considered significant time lag for the signals. The relative exponent $\mathcal{A}$ for the peak $\mathcal{M}_p$ and RMS\begin{figure*}[!t]
	\centering
	\includegraphics[clip]{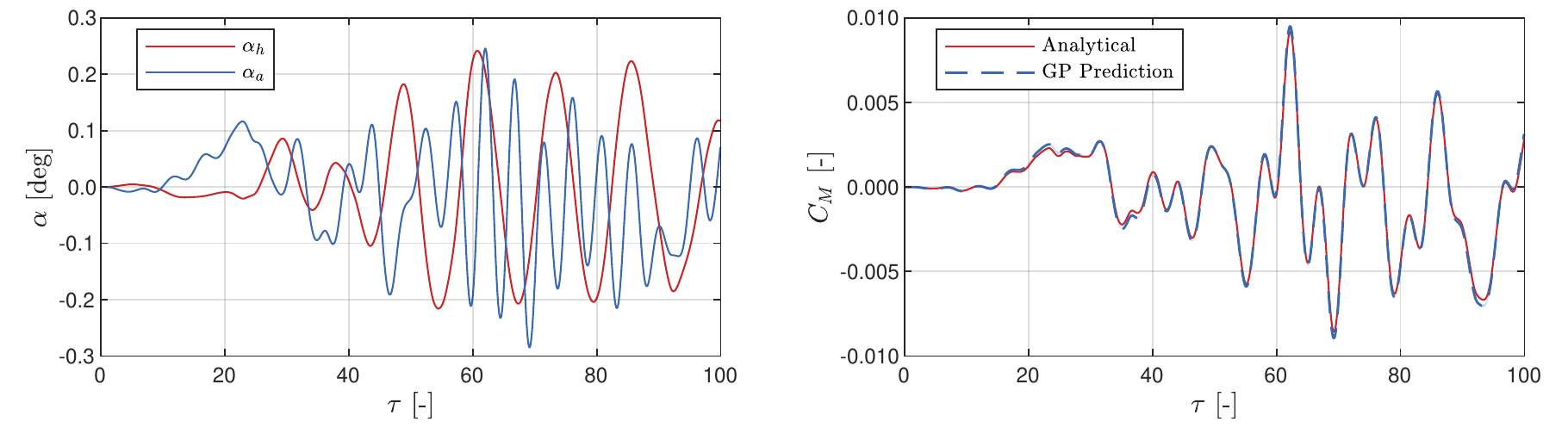} 
	\caption{Flat plate: Prediction for random forced vibration. Random input angles $\alpha_h$ and $\alpha_a$ (left), and corresponding moment coefficient $C_M$ (right). The mean of the GP posterior is represented by the dashed line, while the confidence interval is small (not visible).}
	\label{fig:Ex1a_RandOut}
\end{figure*}
\begin{figure}[!t]
	\centering
	\includegraphics[clip]{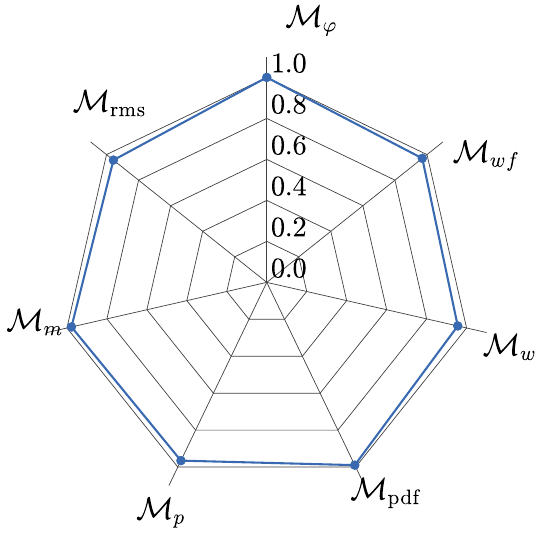} 
	\caption{Flat plate: Comparison metrics $\mathcal{M}_{C_M}^{\mathrm{ANA},\mathrm{GP}}$ for the moment coefficients $C_M$ from random forced vibration for the analytical (ANA) and mean GP models (cf. Fig.~\ref{fig:Ex1a_RandOut}, right). The wavelet central reduced frequency is $K_c$=1.6$\pi$ and the phase normalization time is $T_c$=1/1.6.}
	\label{fig:Ex1a_MetricRand}
\end{figure}
 $\mathcal{M}_{\mathrm{rms}}$ metrics is the relative difference of these quantities. The magnitude metric $\mathcal{M}_{\mathrm{m}}$ represents the difference between the warped time-dependent magnitude of the signals, for which dynamic time warping is used to locally align the signals (by stretching, not scaling). The warping is done to avoid local phase discrepancies (e.g., due to noise) that influence the relative magnitudes. Bhattacharyya distance is utilized as a statistical divergence to formulate the PDF metric $\mathcal{M}_{\mathrm{pdf}}$. The wavelet metric $\mathcal{M}_w$ quantifies the relative discrepancies between the signals on the wavelet time-frequency plane. The Morlet wavelet is used as a mother wavelet with a central frequency $K_c$ that is tuned to provide appropriate resolution in the time-frequency plane according to the signal properties. A similar discrepancy is quantified by the frequency-normalized wavelet metric $\mathcal{M}_{wf}$, only for this metric, the wavelet coefficients are normalized w.r.t. the frequency marginal. If $\mathcal{M}_w$ and $\mathcal{M}_{wf}$ metrics result in similar values, the difference is in the wavelet magnitude rather than the normalized frequency content. For different $\mathcal{M}_{f}$ and $\mathcal{M}_{wf}$, the situation is reversed. The stationarity metric $\mathcal{M}_{s}$ quantifies the difference in the nonstationary part of the wavelet coefficients of the signals. The nonstationary part of the wavelet coefficient is obtained by filtering the stationary part using stationary Fourier surrogates. Finally, the bispectrum metric $\mathcal{M}_{b}$ quantifies the difference in the wavelet bispectrum. With this metric, the discrepancies in the potential second-order nonlinearities (i.e. second-order harmonics) that occur during quadratic-phase coupling can be quantified. \par

\section{Fundamental Application: Flat Plate}\label{sec:FundApp}
This section presents a verification of the presented framework (cf. Fig.~\ref{fig:Framework}) based on the aerodynamic forces and flutter of a flat plate. We use the analytical solution for the linear self-excited forces acting on an infinitely thin flat plate, derived by \cite{Wagner} (time-domain) and \cite{Theodorsen} (frequency-domain). \par 
The aerodynamic forces in the frequency-domain are given in Scanlan's format~\citep{Scanlan1978,Scanlan1990}, in terms of flutter derivatives:
\begin{equation}\label{eq:FP_Forces}
\begin{aligned}
L=&\frac{1}{2}\rho U^2B\left( KH^*_1\frac{\dot{h}}{U}+KH^*_2\frac{B\dot{\alpha}}{U}+K^2H^*_3\alpha+K^2H^*_4\frac{h}{B}\right), \\
M=&\frac{1}{2}\rho U^2B^2\left( KA^*_1\frac{\dot{h}}{U}+KA^*_2\frac{B\dot{\alpha}}{U}+K^2A^*_3\alpha+K^2A^*_4\frac{h}{B}\right),
\end{aligned}
\end{equation}
where $K=2\pi/V_r$ is the reduced frequency, and $H_j^*=H_j^*(K)$ and $A_j^*=A_j^*(K)$ for $j=\{1,\dots,4\}$ are the frequency-dependent flutter derivatives. The analytical expressions that relate the flutter derivatives with the circulatory (analytical) Theodorsen function are given in~\cite{BookScanlan} and are omitted here for the sake of brevity. Equivalently in the time-domain, the aerodynamic forces read \par
\begin{equation}\label{eq:FP_SE_TD}
\begin{aligned}
L&=-\frac{1}{2}\rho U^2 B2\pi \int_{-\infty}^\tau \Phi(\tau-\tau_1)\alpha_e^\prime(\tau_1)\text{d}\tau_1-\frac{ \pi}{4} \rho U^2\left({h}^{\prime\prime}+B\alpha^\prime\right),\\
M&=\frac{1}{2}\rho U^2 B^2\frac{\pi}{2}\int_{-\infty}^\tau \Phi(\tau-\tau_1)\alpha_e^\prime(\tau_1)\text{d}\tau_1-\frac{\pi}{16} \rho B^2U^2\left(\alpha^\prime+\frac{\alpha^{\prime\prime}}{8}\right),\\
\end{aligned}
\end{equation}
where $\Phi=\Phi(\tau)$ is the so-called Wagner function, which asymptotically tends to 1, and it describes the rise-time of the aerodynamic forces. We use the approximation by \cite{JonesRT} for the Wagner function: \mbox{$\Phi$=1-0.16$5\exp$(-0.089$\tau$)-0.335$\exp$(-0.6$\tau$)}. The effective angle $\alpha_e$ is the linearized version of the one in~\eqref{eq:EffAngle} (i.e. the $\arctan$ is dropped), with aerodynamic center $m$=0.25. \par  
The numerical parameters for learning and prediction (forced and free vibration) of the aerodynamic forces are given in Tab.~\ref{tab:NumParam}, along with the parameters for the two bridge decks studied in Sec.~\ref{sec:BridgeAero}. In this case, the selected number of lags and time step ($S$=200, $\Delta\tau$=0.05) correspond to fading memory time of $\tau$=10. For comparison, the unit-impulse response function, which corresponds to the Wagner function, reduces to 3\% at $\tau$=10. The attenuation time of the unit-impulse response function can be related to the number of lag terms, at least in a linear sense. 

\begin{figure*}[!t]
	\centering
	\includegraphics[clip,width=\textwidth]{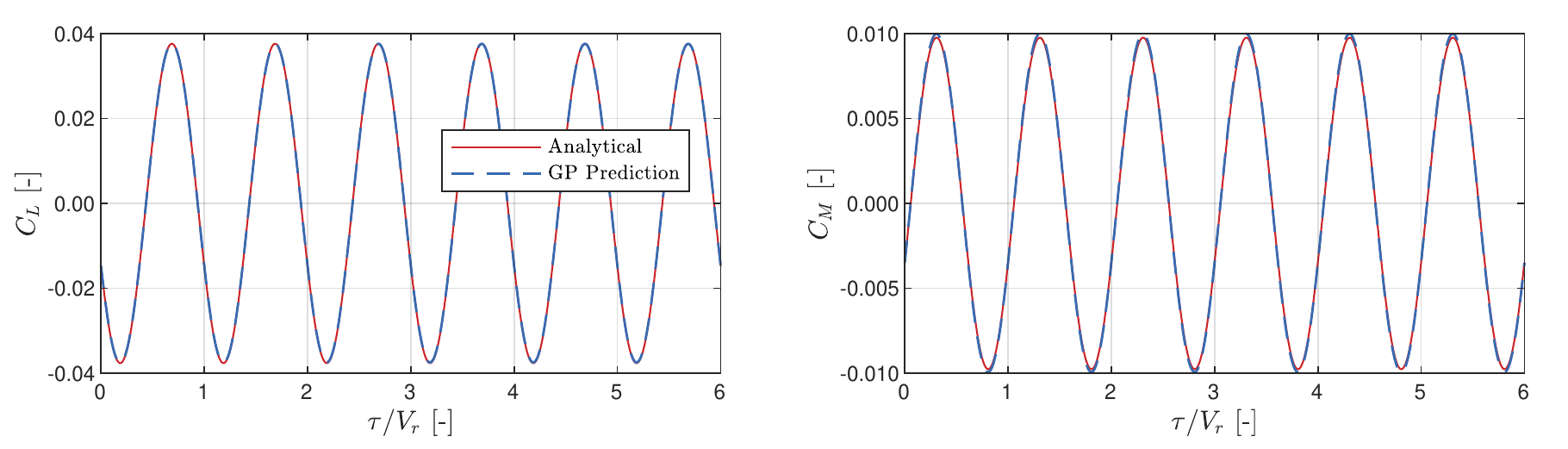} 
	\caption{Flat plate: Prediction for sinusoidal forced rotation. Lift $C_L$ (left) and moment $C_M$ (right) coefficients. The mean of the GP posterior is represented by the dashed line, while the confidence interval is small (not visible). The reduced velocity of the oscillation is $V_r$=6 with an amplitude of $\alpha_0$=0.1 deg.}
	\label{fig:Ex1a_HarmonicOut}
\end{figure*}

\begin{figure}[!t]
	\centering
	\includegraphics[clip]{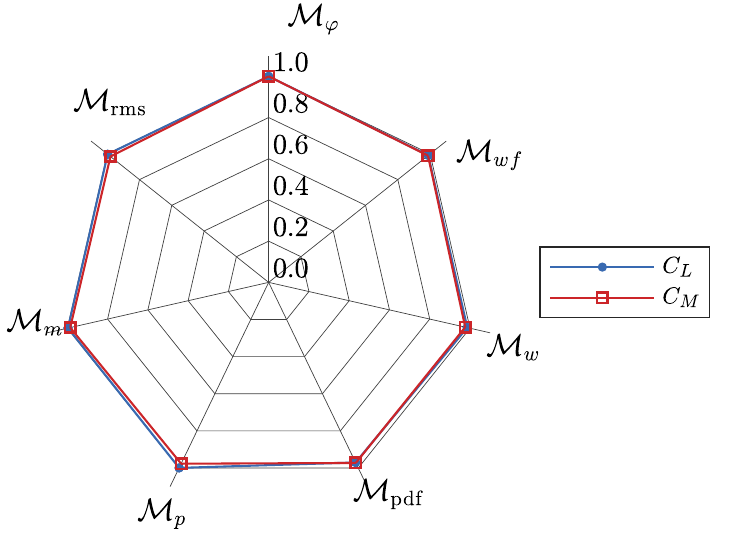} 
	\caption{Flat plate: Comparison metrics $\mathcal{M}^{\mathrm{ANA},\mathrm{GP}}$ for the wind coefficients $C_L$ and $C_M$ from sinusoidal forced rotation for the analytical (ANA) and mean GP models (cf. Fig.~\ref{fig:Ex1a_HarmonicOut}). The wavelet central reduced frequency is $K_c$=12$\pi$ and the phase normalization time is $T_c$=1/12.}
	\label{fig:Ex1a_MetricSin}
\end{figure}
\subsection{Learning}
Figure~\ref{fig:Ex1a_Train} (top) depicts the generated training signals, while Fig.~\ref{fig:Ex1a_Train} (bottom) include a sample of the wind coefficients for the analytical model and the mean GP prediction for the training input data-set. This sample of $C_L$ and $C_M$ corresponds to the gray rectangle in Fig.~\ref{fig:Ex1a_Train} (top).  A small signal-to-noise (SNR) ratio is added to the analytical wind coefficient with SNR=20 to eliminate numerical issues, which arise in the covariance matrix inversion during the training procedure. Thus, the output could be generally considered noiseless. Therefore, for the learned hyperparameters, the mean of the GP posterior fits well for the training input motion with minor variance.
\begin{figure*}[!t]
	\centering
	\includegraphics[clip]{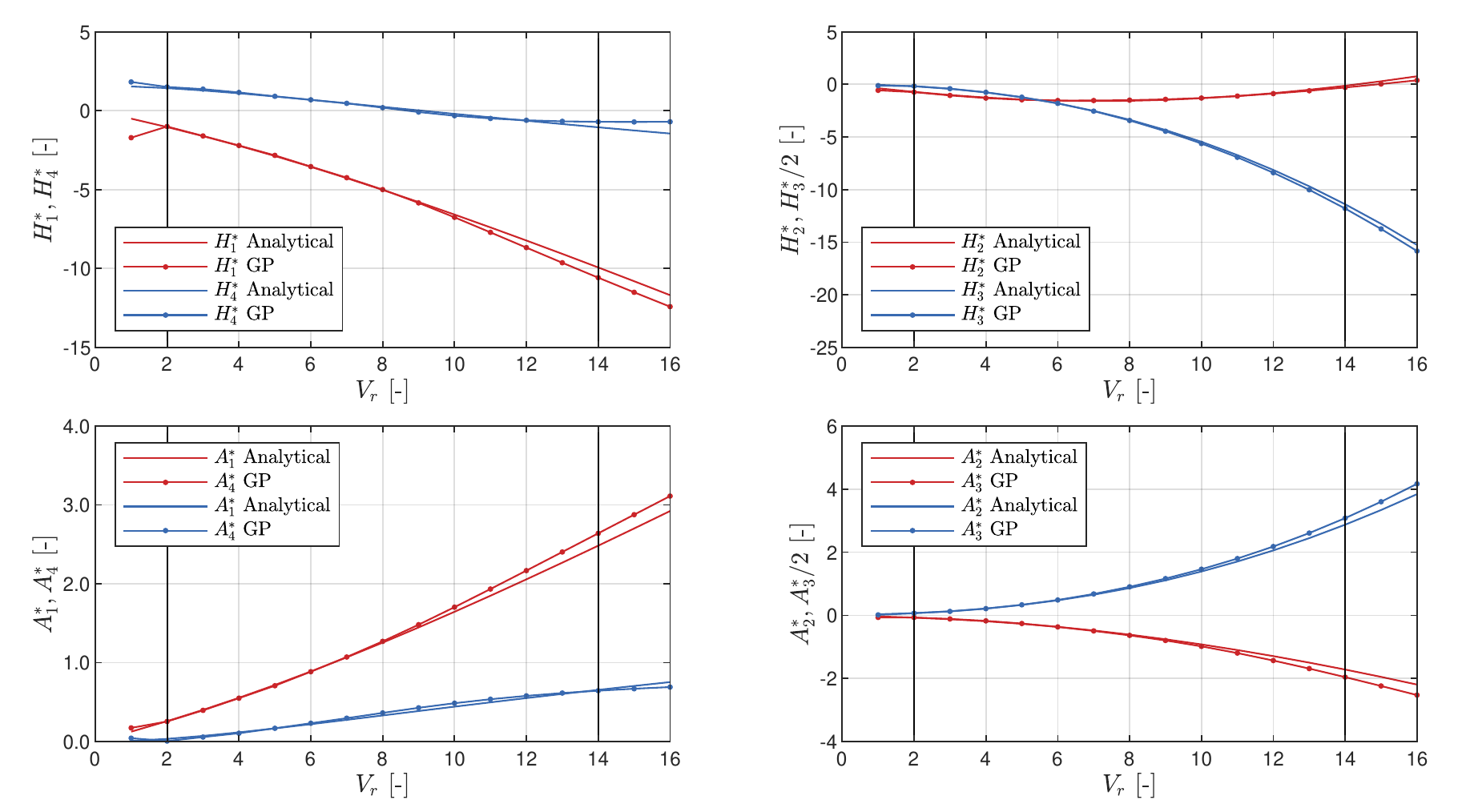} 
	\caption{Flat plate: Prediction of the flutter derivatives for sinusoidal oscillations with amplitudes of $\alpha_0$=0.1 deg. The GP prediction is based on the mean of the posterior. The black lines represents the end of the training range for the reduced velocity $V_r$.}
	\label{fig:Ex1a_FlutterDer}
\end{figure*}
\begin{figure}[!t]
	\centering
	\includegraphics[clip]{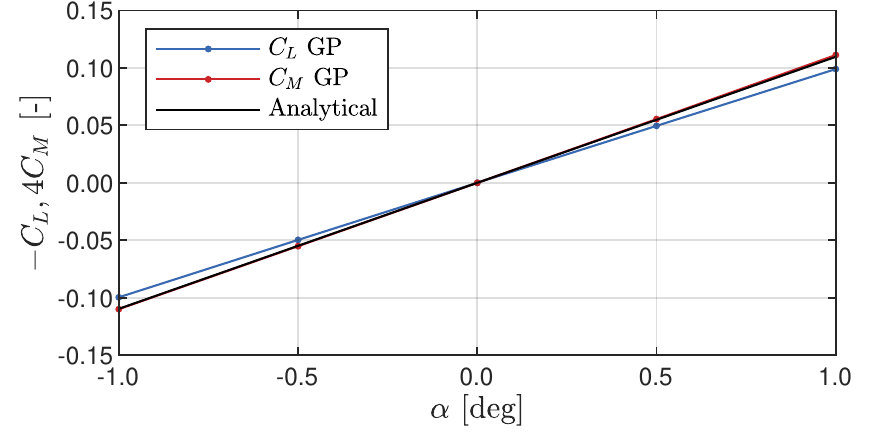} 
	\caption{Flat plate: Prediction of the static wind coefficients. The GP prediction is based on the mean of the posterior. The analytical solution is computed for a flat plate as: $-C_L=4C_M=2\pi\alpha$.}
	\label{fig:Ex1a_StatWind}
\end{figure}
\subsection{Prediction: Forced vibration}
The trained GP model is tested here for an input motion that is different from the training data in a multi-step ahead fashion; i.e. the entire input motion is fed at once (cf. Sec.~\ref{sec:FrameWork}, step iv). To assess the mean aerodynamic force for the GP model, we use seven of the nine comparison metrics discussed in Sec.~\ref{sec:FrameWork} and take the results of the analytical model as a reference. The stationarity $\mathcal{M}_s$ and bicoherence $\mathcal{M}_b$ are not considered since they are obsolete for a linear model. For the discussion in this section, only the mean of the GP predictive posterior is considered since the variance is insignificant due to the low noise. \par
Initially, the plate is forced to oscillate with a random coupled motion. The input motion is taken from a plate response during 2D buffeting analysis~\citep{PhDKavrakov}, and it is scaled such that it is in the appropriate amplitude range. The signal is broadband with the main frequency of the vertical displacements corresponding to $V_r\approx$6 and rotational frequency corresponding to $V_r\approx$13.  The input signals are shown in Fig.~\ref{fig:Ex1a_RandOut} (left), including the prediction of the moment coefficient $C_M$ for the analytical and GP models (right). Excellent correspondence can be observed, with a low variance of the GP model such that it is not visible in the figure. The comparison metrics are shown in  Fig.~\ref{fig:Ex1a_MetricRand}. The metrics support the previous conclusion as their values are $\mathcal{M}\geq$0.95 for all the signal features captured by them. The presented results for random forced vibration show the framework's potential to be used in a buffeting analysis.\par
A convenient way to verify the linear fluid memory is by comparing the flutter derivatives, which are determined based on sinusoidal forced vibration. The deck is forced to oscillate sinusoidally for a range of reduced velocities (cf. Tab.~\ref{tab:NumParam}). For example, the predictions of both models' wind coefficients are compared in Fig.~\ref{fig:Ex1a_HarmonicOut} for $V_r$=6 (the mean, in case of the GP model), indicating excellent correspondence. Additionally, the comparison metrics for these time-histories (cf. Fig.~\ref{fig:Ex1a_MetricSin}) yield a similar conclusion for all signal features. The flutter derivatives are computed for the selected frequency range and beyond. These are shown in Fig.~\ref{fig:Ex1a_FlutterDer}, where the black lines indicate the start and end of the training frequency range. In the GP model case, the flutter derivatives are based on the mean of the predictive posterior. Excellent correspondence can be noted for the low reduced velocities in the training range. The flutter derivatives branches are slightly diverging from one another for higher reduced velocities. Beyond the training range, the flutter derivatives diverge but remain stable for high reduced velocities $V_r>14$. Outside the training range at $V_r=1$, a difference is noted in $H_1^*$; however, it is noted that the prediction was still purely sinusoidal, i.e. no numerical instability was observed. \par
Moreover, Fig.~\ref{fig:Ex1a_StatWind} shows the prediction of the static wind coefficients of the GP model compared to analytical solution ($-C_L=4C_M=2\pi\alpha$). The static wind coefficients can be regarded as the asymptotic values to the flutter derivatives at $V_r\rightarrow\infty$~\citep{Diana2013}. The input effective angle for the GP model only contains values for $\alpha_a$ for the whole memory, while the rest is zero. Excellent correspondence can be observed for the moment and minor deviation for the lift. The prediction of the static wind coefficients and flutter derivatives outside the training range show that the GP model yields fair and, more importantly, stable results during extrapolation in the frequency plane, i.e. an overfit is avoided (cf. Sec.~\ref{sec:Learning}). However, it is difficult to guarantee model robustness outside the training space (amplitude or frequency). This also depends on the numerical parameters such as time step and number of lags. Therefore,  as with any data-driven model, the training input should be carefully designed to cater for the frequencies and amplitudes of interest.

\begin{figure*}[!t]
	\centering
	\includegraphics[clip]{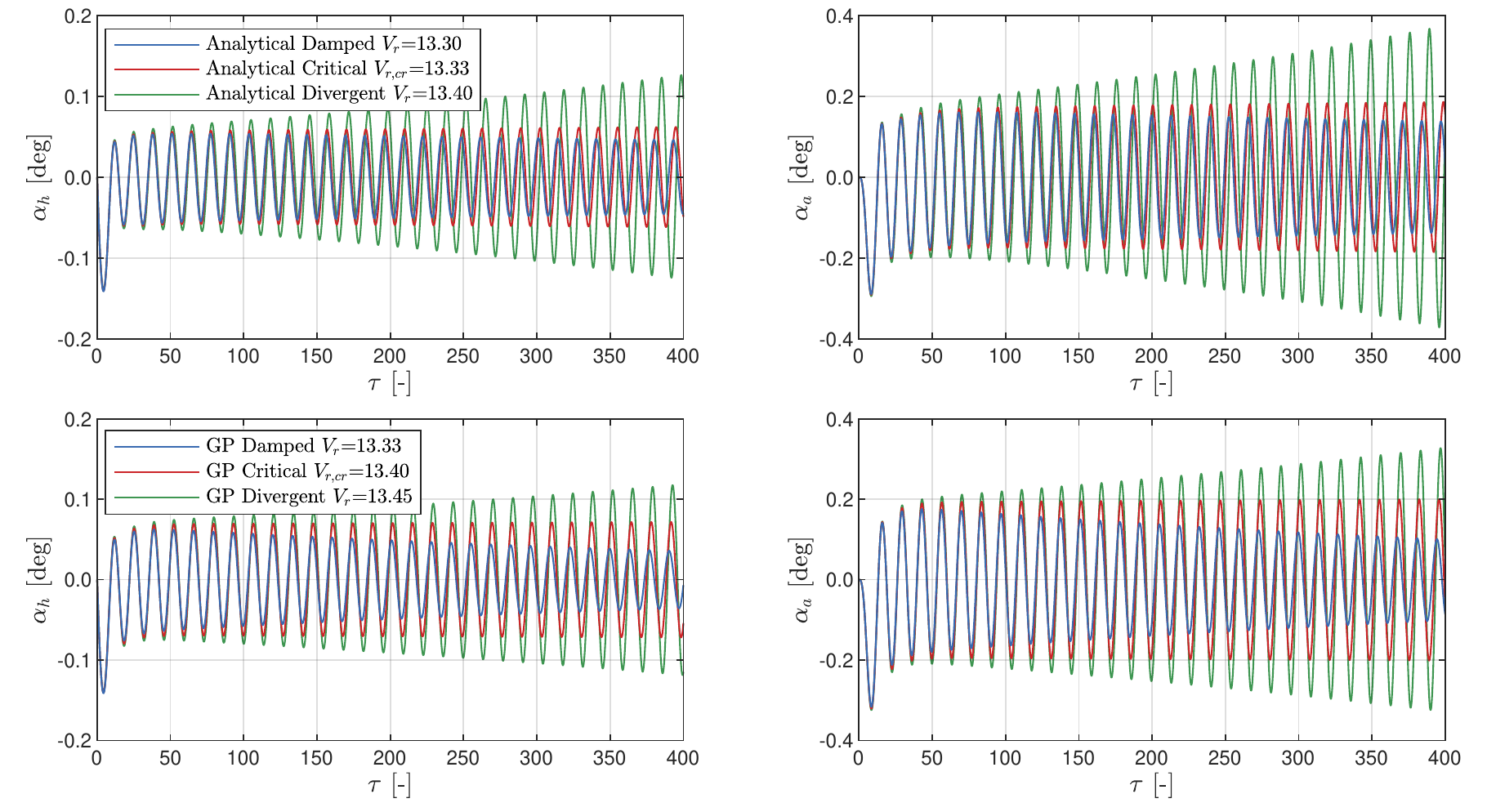} 
	\caption{Flat plate: Flutter prediction from free vibration analyses with prescribed initial displacements for the analytical model (top) and the GP model (bottom). The reduced wind speed is computed as $V_r=U/(f_{h\alpha}B)$, where $f_{h\alpha}$ is the mean of the vertical and torsional frequencies.}
	\label{fig:Ex1b_Flut}
\end{figure*}
\begin{figure*}[!t]
	\centering
	\includegraphics[clip,scale=1]{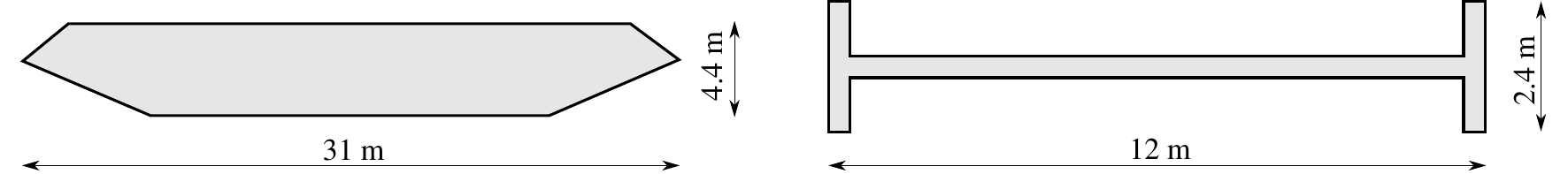} 
	\caption{Schematic of the two studied bridge decks: Great Belt's (left) and H-shaped deck (right).}
	\label{fig:Schematic_TCGB}
\end{figure*}
\begin{figure*}[!t]
	\centering
	\hspace*{-4cm}\includegraphics[clip,scale=1.5]{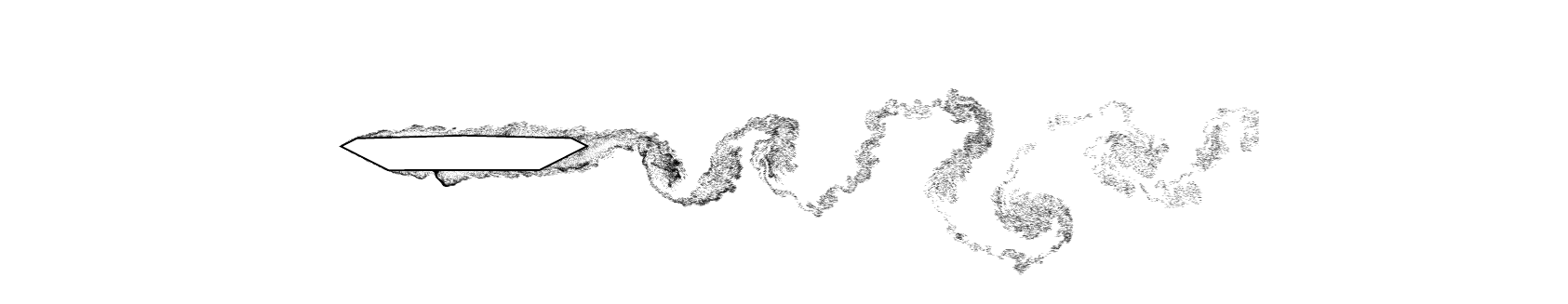} 
	\caption{Great Belt Bridge: Instantaneous particle map from a CFD simulation.}
	\label{fig:GB_SE_ParticleMap}
\end{figure*}

\subsection{Prediction: Flutter}\label{sec:FPFluter}
Finally, the coupled flutter response is computed for a 2D coupled model (cf.~\eqref{eq:EqMot}). The structural parameters of the flat plate are given in Tab.~\ref{tab:NumParam}.
 The standard Newmark-Beta algorithm is used for the integration in time~\citep{BookClough}. In the case of the GP model, the analysis is conducted by feeding the mean of the predictive posterior in a step-by-step ahead fashion (cf. Sec.~\ref{sec:FrameWork}, step iv)). The propagation of uncertainty is not considered. \par
The deck is subjected to initial displacements and let to oscillate freely for several wind speeds. Figure~\ref{fig:Ex1b_Flut} depicts the time-histories of both DOFs during damped, critical, and divergent oscillations. For both models, the flutter is coupled, and similar behavior is observed in all three oscillation regimes. The critical reduced velocity for the analytical model is $V_{r,cr}=U/(f_{h\alpha}B)$=13.33 and for the GP model is $V_{r,cr}=U/(f_{h\alpha}B)$=13.40, where $f_{ha}=(f_a+f_h)/2$, which yields a difference of 0.5\% that can be considered as insignificant. Obtaining a coupled flutter with similar critical velocities concludes the verification of the data-driven GP model for linear aerodynamics.
\begin{figure*}[!t]
	\centering
	\includegraphics[clip]{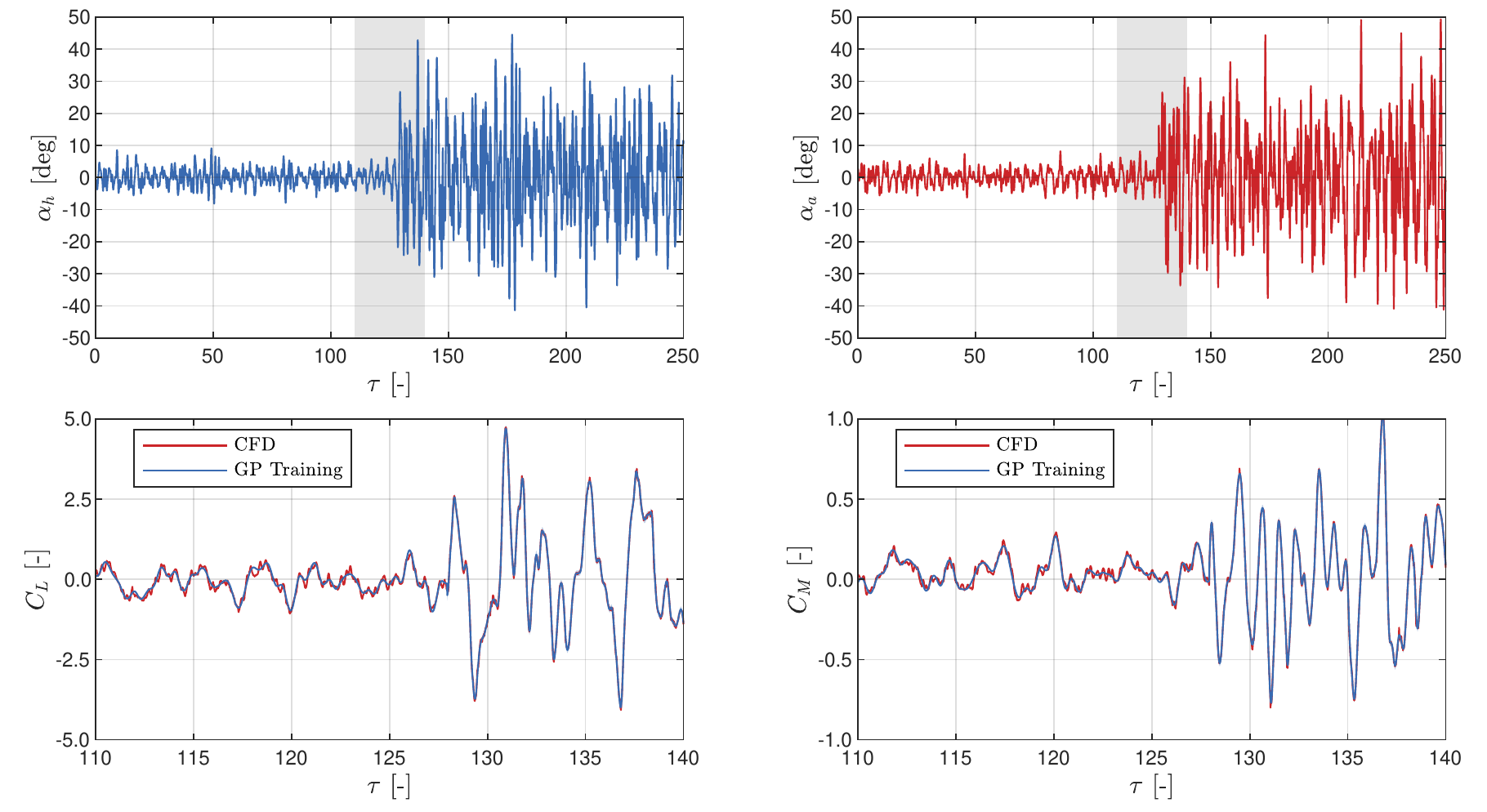} 
	\caption{Great Belt Bridge: Training. Input random excitation for both degrees of freedom $\alpha_h$ and $\alpha_a$ (top) and output wind coefficients $C_L$ and $C_M$ (bottom). The wind coefficients (bottom) are depicted only for the corresponding shaded area of the input (top). The mean of the GP posterior is depicted with a solid line, while the 99\% confidence interval is small, and thus, not visible. The training is two amplitudes with $\sigma_{\alpha_a}=\sigma_{\alpha_h}$=2.5 deg and $\sigma_{\alpha_a}=\sigma_{\alpha_h}$=15 deg. The reduced velocity range is $V_r$=2-16.}
	\label{fig:Ex2a_Train}
\end{figure*}

\section{Application to Bridge Aerodynamics}\label{sec:BridgeAero}
Next, we apply the framework to model the forces acting on two bridge decks, which arguably present benchmarks in bridge aerodynamics: a simplified H-shapedd deck and Great Belt's streamlined section (cf. Fig.~\ref{fig:Schematic_TCGB}). The aerodynamic GP model is first trained based on CFD simulation and then the predictions from forced and free vibration (flutter) for both models are compared (cf.~Fig.~\ref{fig:Framework}). The CFD model is considered as a reference for the comparison.\par
The fluid-structure interaction is simulated utilizing the 2D Vortex Particle Method as a CFD method to solve the governing fluid equations numerically. The vortex particle method is a grid-free Lagrangian method, which discretizes the vorticity transport form of the Navier-Stokes equations using vortex particles that carry concentrated circulation. For an immersed body, the boundary conditions are enforced through Biot-Savart law, which results in the circulation on the boundary, and virtually, aerodynamic forces. The boundary element method is used for discretizing the body on panels. The vortex methods enjoy an efficient implementation in 2D and has shown success, particularly for unresolved high Reynolds number-type applications, such as bluff body aerodynamics.  Without going into details, we refer to previous works on the formulation~\citep{BookCottet}, present numerical implementation~\citep{MorgenthalMCRobie,MorgenthalWalther,MorgenthalGPU}, and applications to bridge aerodynamics~\citep{Larsen1998,GeXiang,Farsani,McRobie2013,KavrakovArgentini}. We note that using the Vortex Particle Method to simulate the wind-structure interaction is merely for proof of concept. Any other CFD method or experiments can be used for the same purpose.\par
\subsection{Great Belt Bridge}
The deck of the Great Belt Bridge is the one of the most studied sections in bridge aerodynamics. It is a single box girder with a chord of 31 m and an aspect ratio of 7 (cf. Fig.~\ref{fig:Schematic_TCGB}, left). A simplified version of the deck, section H4.1 from~\cite{Reinhold}, is used in the CFD modeling, which does not include any auxiliary equipment (e.g., cables, railings, barriers). The simulations are conducted with a constant Reynolds number and reduced time step (cf. Tab.~\ref{tab:NumParam}, for discretization of the section on 250 panels. For illustration, Fig.~\ref{fig:GB_SE_ParticleMap} depicts an instantaneous particle map from a CFD simulation. Since we use similar numerical parameters and some of the results from~\cite{KavrakovMorgenthal2}, we refer to it for further details on the CFD modeling, which also includes comparison/validation with other studies.\par 
Parameters for constructing and training the GP-NFIR model are given in~Tab.~\ref{tab:NumParam} (identical for lift and moment). With the selected number of lag terms and time step ($\tau_{\mathrm{GP}}=0.05$,$S=200$), the GP model captures the fading fluid memory up to $\tau$=10 time units in the past, which is similar to the lag time for the flat plate. There is not yet a consistent method to select the number of terms for bluff bodies. \cite{AbbasANN} used up to 3 terms; however, their NFIR model included the velocity and acceleration as a lagged input. \cite{LiWu} used equivalent lags up to $\tau=20$. Although $S$=200 proved to be sufficient for the present study, the ratio of time step vs. the number of lags warrants a further investigation.

\begin{figure*}[!t]
	\centering
	\includegraphics[clip]{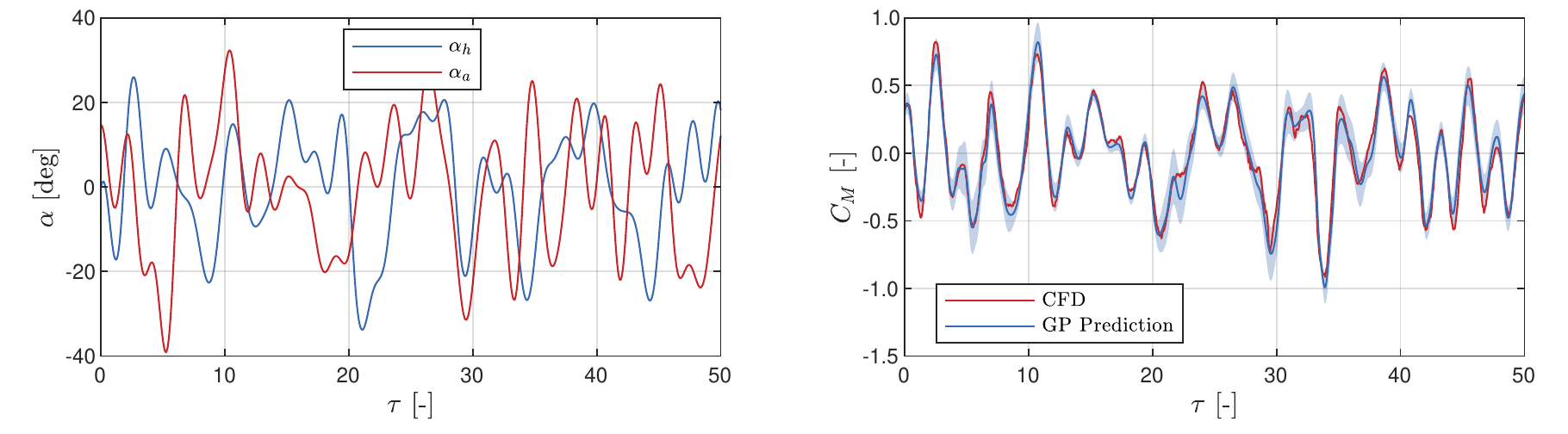} 
	\caption{Great Belt Bridge: Prediction for random forced vibration ($2\leq V_r\leq16$). Random input angles $\alpha_h$ and $\alpha_a$ (left), and corresponding moment coefficient $C_M$ (right). The mean of the GP posterior is represented by the solid line, while the shaded area represents the 99\% confidence interval.}
	\label{fig:Ex2a_Rand}
\end{figure*}

\begin{figure}[!t]
	\centering
	\includegraphics[clip]{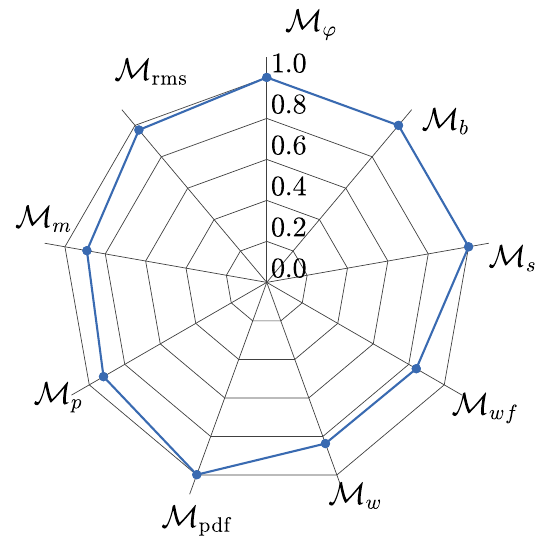} 
	\caption{Great Belt Bridge: Comparison metrics $\mathcal{M}_{C_M}^{\mathrm{CFD},\mathrm{GP}}$ for the moment coefficients $C_M$ from random forced vibration for the CFD and mean GP models (cf. Fig.~\ref{fig:Ex2a_Rand}, right). The wavelet central reduced frequency is $K_c$=2$\pi$ and the phase normalization time is $T_c$=8.}
	\label{fig:Ex2a_MetricRand}
\end{figure}
\subsubsection{Learning}
Figure~\ref{fig:Ex2a_Train} (top) depicts the input training signals generated based on the parameters in Tab.~\ref{tab:NumParam}, with two amplitudes stacked together. A sample of the resulting output lift $C_L$ and moment $C_M$ coefficients are shown at the bottom of the same figure (GP mean prediction for training input and CFD as target; the sample corresponds to the shaded area of the input signals). The mean of the GP posterior for the training input results in a good fit for the CFD data for both amplitude ranges, with a narrow confidence interval that is not visible in the figure. An additional high-frequency component for the CFD model can be noted in the low amplitude range due to the so-called interior noise (i.e. vortex shedding and local separation). This effect cannot be accounted for using the presented model as it requires additional input, i.e. there are forces due to vortex shedding even for a stationary deck. However, the interior noise forces are low compared to the motion-induced forces for this deck, even for the low amplitude range. The idea of selecting two amplitude ranges is to capture the aerodynamics for low and high amplitude oscillation (i.e. the linear and nonlinear range). It is challenging to select the input amplitudes for which the system can be assumed linear. In this case, it is selected as to $\sigma_{\alpha1}$=2.5 deg as the static wind coefficients are generally linear in the $\pm$5 deg range (not shown, cf.~\cite{KavrakovMorgenthal2}).\par

\begin{figure*}[!t]
	\centering
	\includegraphics[clip]{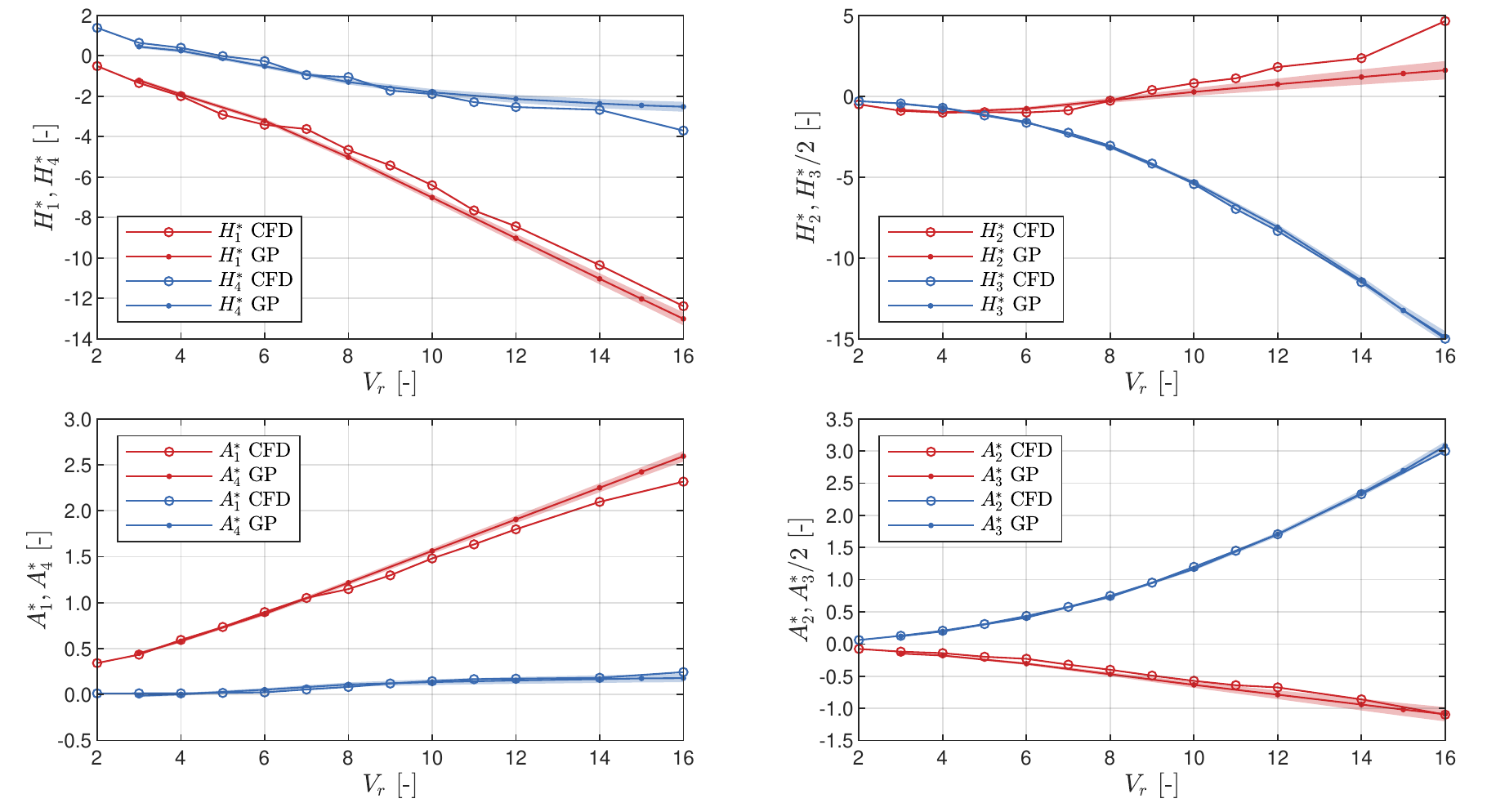} 
	\caption{Great Belt Bridge: Prediction of the flutter derivatives for forced sinusoidal motion with an amplitude of $\alpha_0$=1 deg, around zero static angle of attack. The mean (solid line) and 99\% confidence interval (shaded area) for the GP model are obtained from 1000 samples that are drawn from its predictive posterior at each $V_r$.}
	\label{fig:Ex2a_FlutterDer}
\end{figure*}

\begin{figure*}[!t]
	\centering
	\includegraphics[clip]{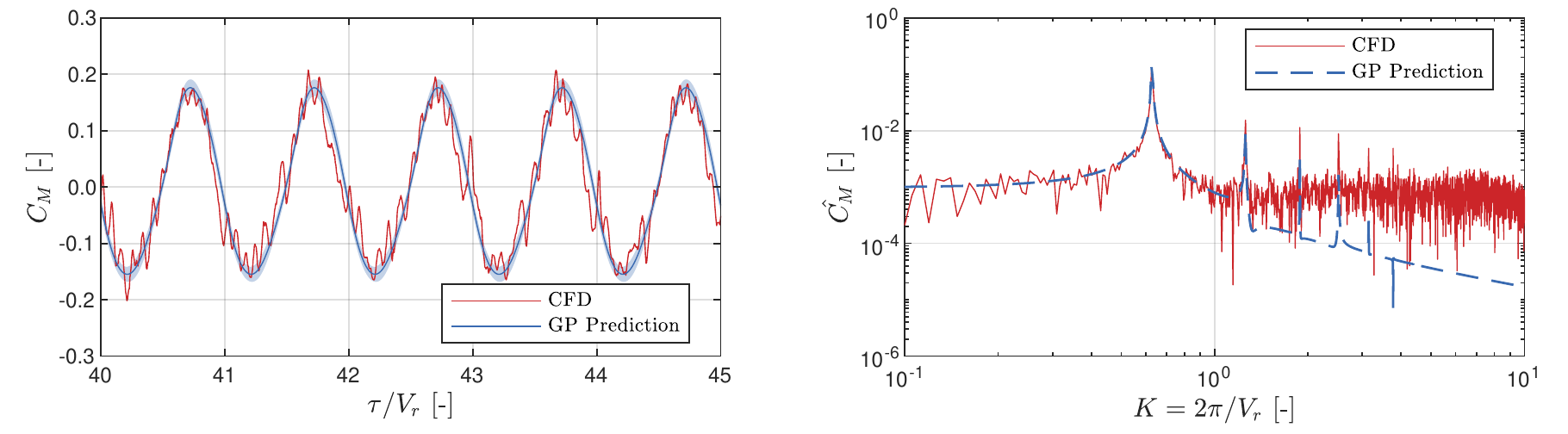} 
	\caption{Great Belt Bridge: Prediction for forced sinusoidal rotation with a large amplitude ($\alpha_0$=15 deg, $V_r$=10). A sample of 5 cycles of the moment coefficient time-history ${C}_M$ is shown (left), and the corresponding FFT $\hat{C}_M$ (right). The mean of the GP posterior is represented by the solid line, while the shaded area represents the 99\% confidence interval. The FFT for the GP model is based on the mean.}
	\label{fig:Ex2a_Sine}
\end{figure*}
\subsubsection{Prediction: Forced vibration}
The prediction quality of the trained GP model is assessed by subjecting the deck to forced vibration. The GP predictive distribution is obtained in a multi-step ahead manner for the entire input motion (cf. Sec.~\ref{sec:FrameWork}, step iv)).\par 
Initially, we examine the aerodynamic forces for a combination of random torsional and vertical motion, with similar frequency and amplitude ranges as for the high-amplitude training signal (cf. Tab.~\ref{tab:NumParam}). Figure~\ref{fig:Ex2a_Rand} depicts the input motion (left) and the moment coefficient $C_M$ as the output (right). The shaded area in the moment coefficient is the 99\% confidence interval of the GP model's predictive posterior. Good correspondence can be seen, with the GP posterior distribution enclosing the target CFD results. Figure~\ref{fig:Ex2a_MetricRand} depicts the comparison metrics for the mean prediction of the GP model and the CFD prediction, taking the CFD as a reference. Most of the metrics result in relatively high values, indicating good correspondence. The bispectrum metric amounts to $\mathcal{M}_b^{\mathrm{GP,CFD}}$=1 since no second-order nonlinearity (i.e. quadratic-phase coupling) is identified in the signals. However, this does not mean that the moment coefficient is linear in this range. Rather, the resulting nonlinearity is not a second-order nonlinearity. Aerodynamic forces at such high angles of attack experience nonlinearity due to flow separation at the leading edge even for streamlined bridge decks, as it is concluded by~\cite{KavrakovMorgenthal2} for this particular case by comparing the CFD with a linear semi-analytical model. The angle amplitudes are case-dependent depending on the perceived aeroelastic oscillations (e.g. wind speed, deck type, dynamic properties), which, in turn, determines whether the forces will be linear. High angle amplitudes were selected here to cover the LCO range (as seen later), and initiate nonlinear behavior. Future studies may suggest recommendations of typical oscillation amplitudes for training (e.g. during buffeting), which are easily definable since the presented model is nondimensional. \par
Next, the linear range is assessed through the flutter derivatives around zero static angle of attack. The section is forced to perform sinusoidal oscillations for a selected reduced velocity range and forcing amplitude (cf. Tab.~\ref{tab:NumParam}). The mean and the confidence interval of the GP model's flutter derivatives are identified by sampling from its predictive posterior. Figure~\ref{fig:Ex2a_FlutterDer} depicts the flutter derivatives for the CFD and the GP model. The mean and the 99\% confidence interval for the GP models are identified based on 1000 samples. The uncertainty is generally low, and good correspondence is obtained between the GP-based and CFD-based derivatives. The confidence interval for the GP model does not enclose the CFD-based derivatives in some instances. Although it cannot be pinpointed exactly what is the reason for this, possibles reasons include: the CFD training data set did not include sufficient information on the derivatives, high noise during CFD determination of the derivatives, or that the GP was not setup up/trained adequately and thus the noise $\sigma_{nL},\sigma_{nM}$ was underestimated. It is worth noting that the CFD-based derivatives used here are successfully validated in~\cite{KavrakovMorgenthal2} with experiments conducted by~\cite{Reinhold}. \par
\begin{figure}[!t]
	\centering
	\includegraphics[clip]{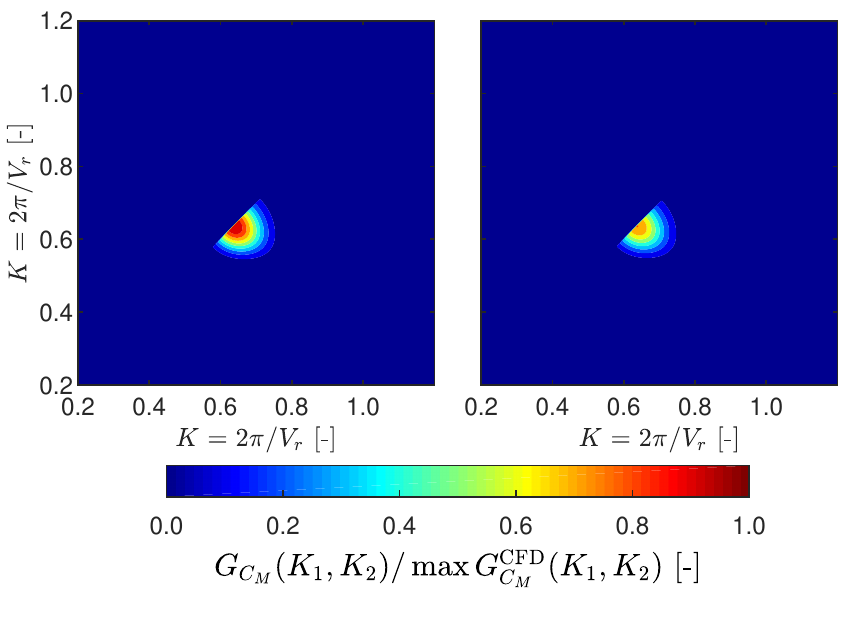} 
	\caption{Great Belt Bridge: Normalized filtered bispectrum of the moment coefficient for the CFD $G_{C_M}^\mathrm{CFD}$ (left) and GP models $G_{C_M}^\mathrm{GP}$ (right) for forced sinusoidal rotation with a large amplitude ($\alpha_0$=15 deg, $V_r$=10).}
	\label{fig:Ex2a_BiSpec}
\end{figure}
\begin{figure}[!t]
	\centering
	\includegraphics[clip]{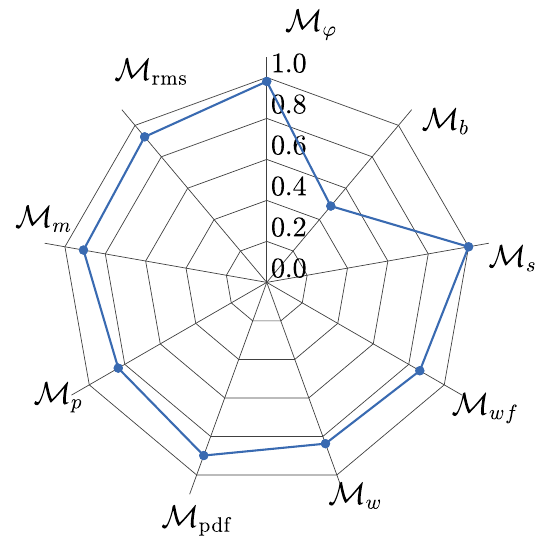} 
	\caption{Great Belt Bridge: Comparison metrics $\mathcal{M}_{C_M}^{\mathrm{CFD},\mathrm{GP}}$ for the moment coefficients $C_M$ from forced sinusoidal rotation with high amplitudes for the CFD and mean GP models (cf. Fig.~\ref{fig:Ex2a_Sine}, left).}
	\label{fig:Ex2a_MetricSineHighAmp}
\end{figure}

Lastly, we compare the models for a high-amplitude sinusoidal input motion to excite and observe aerodynamic nonlinearities in the output forces. The section rotates with an amplitude of $\alpha_0$=15 deg with a reduced velocity of $V_r$=10. Figure~\ref{fig:Ex2a_Sine} (left) depicts the mean and confidence interval of the resulting moment $C_M$ for the GP model, which is compared to the CFD prediction. The corresponding fast Fourier transform (FFT) $\hat{C}_M$ is shown in the same figure (right). In the case of the GP model, $\hat{C}_M$ is based on the mean of the predictive posterior of $C_M$. The $C_M$ for the GP model preserves the higher-order harmonics to a certain extent, while filtering the vortex-shedding and interior noise. Figure~\ref{fig:Ex2a_BiSpec} presents the filtered bispectrum of the moment $G_{C_M}$ for the CFD model and for the mean of the GP posterior. The filtered bispectrum of the moment $G_{C_M}$ for both models exhibit a distinctive peak at the frequency couple $(K_1,K_2)$=(2$\pi$/10,2$\pi$/10), which corresponds to the input motion frequency. The bispectrum is filtered using surrogates to ensure that this peak and the second-order harmonics appearing in Fig.~\ref{fig:Ex2a_Sine} (right) are due to nonlinear interaction, i.e. that quadratic phase coupling occurs in the $C_M$ (cf.~\cite{KavrakovKareem} for details on the filtering procedure). The GP model underestimates the amplitude of second-order harmonic in $C_M$. The bispectrum metric (cf. Fig.~\ref{fig:Ex2a_MetricSineHighAmp}) resulted in $\mathcal{M}_{C_M}^{\mathrm{CFD},\mathrm{GP}}\approx$0.5, although it may be argued that the bispectrum metric tends to be quite sensitive. The rest of the metrics attain high values. This shows that the GP model is capable of capturing nonlinearities in terms of higher-order harmonics to a certain extent. The contribution of the higher-order harmonics is relatively low compared to the first harmonic. Further studies including experimental validation for complicated bridge decks are warranted. Moreover, higher-order harmonics are usually relatively higher for the drag forces~\citep{Skyvulstad2021}, which would be a natural extension of the present framework.

\begin{figure*}[!t]
	\centering
	\includegraphics[clip]{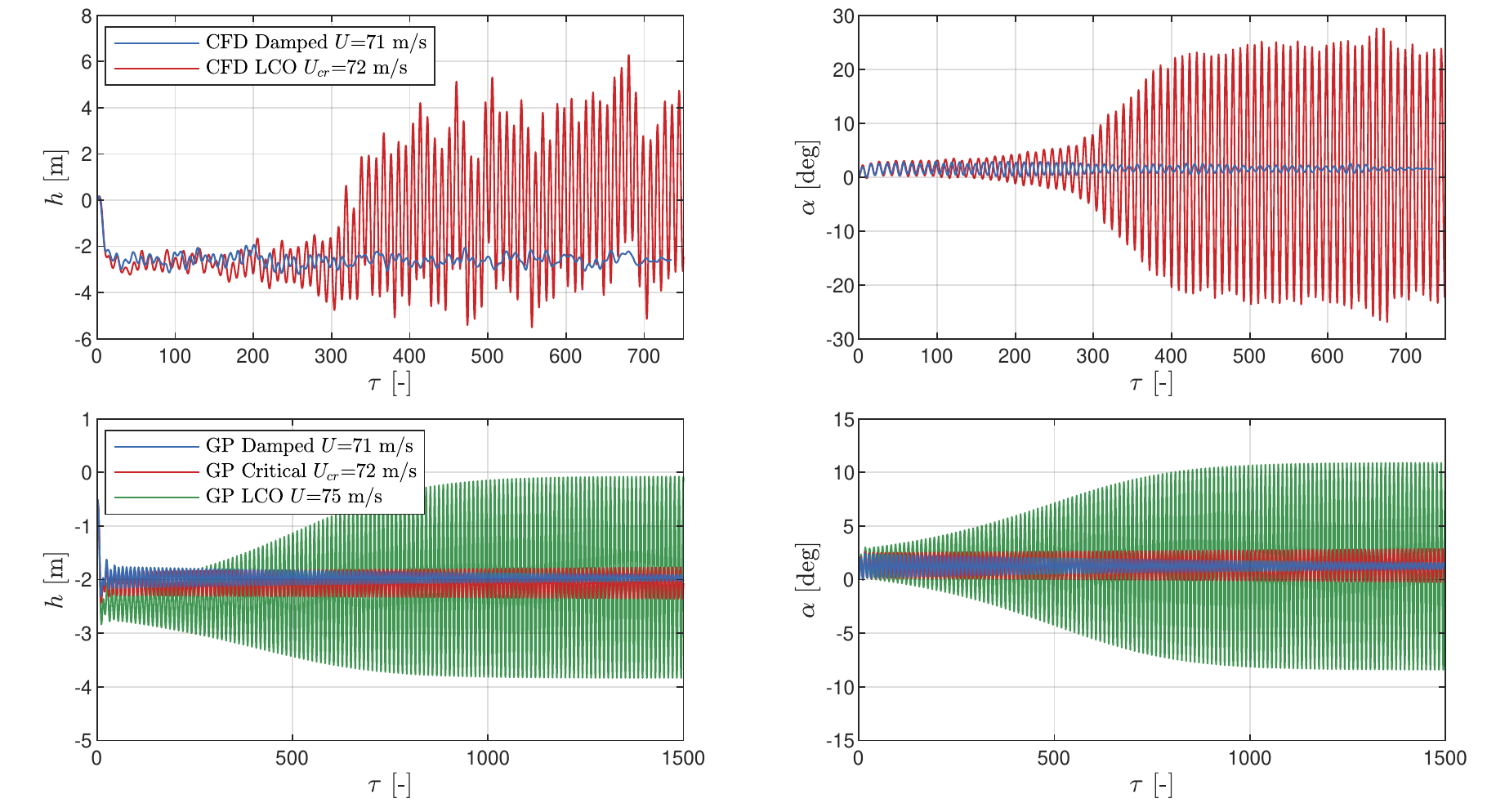} 
	\caption{Great Belt Bridge: Flutter and LCO prediction from free vibration analyses with prescribed initial displacements for the CFD model (top) and the GP model (bottom).}
	\label{fig:Ex2b_Flut}
\end{figure*}

\begin{figure*}[!t]
	\centering
	\includegraphics[clip]{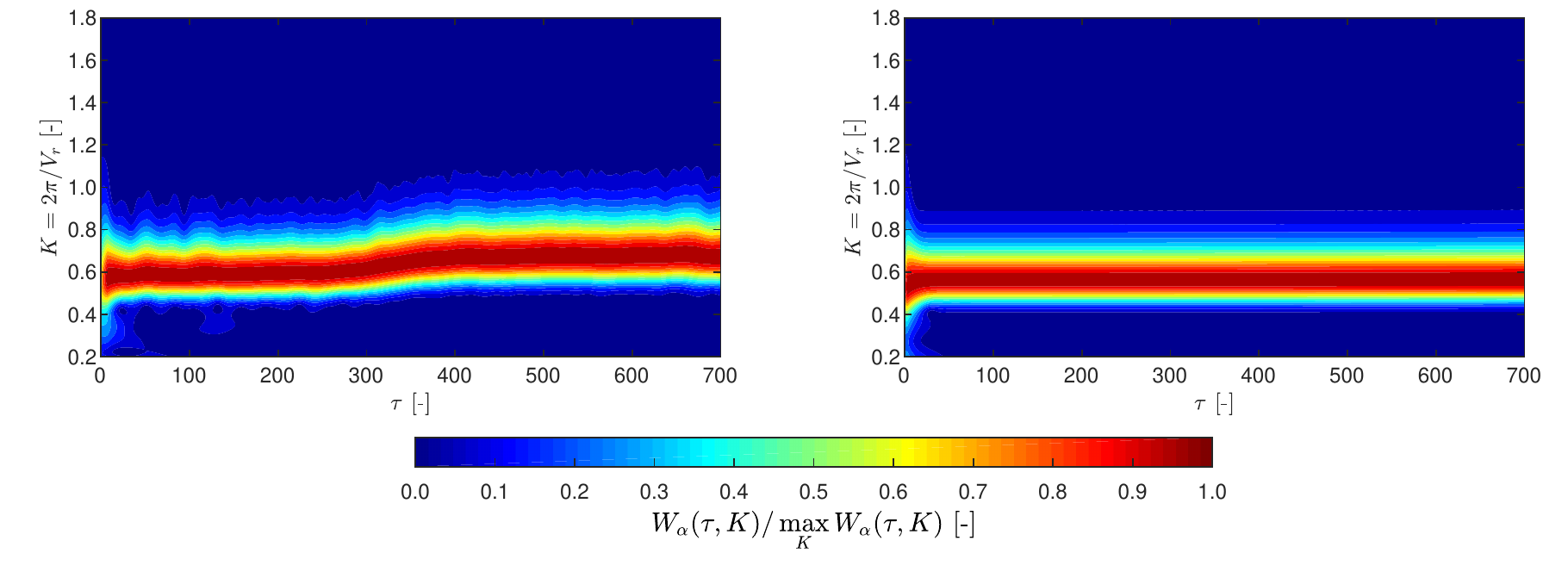} 
	\caption{Great Belt Bridge: Frequency-normalized wavelet magnitude of the rotation $\alpha$ during flutter for the CFD (left) and GP model (right) at $U$=72 m/s.}
	\label{fig:Ex2b_WaveletFlut}
\end{figure*}

\begin{figure*}[!t]
	\centering
	\includegraphics[clip]{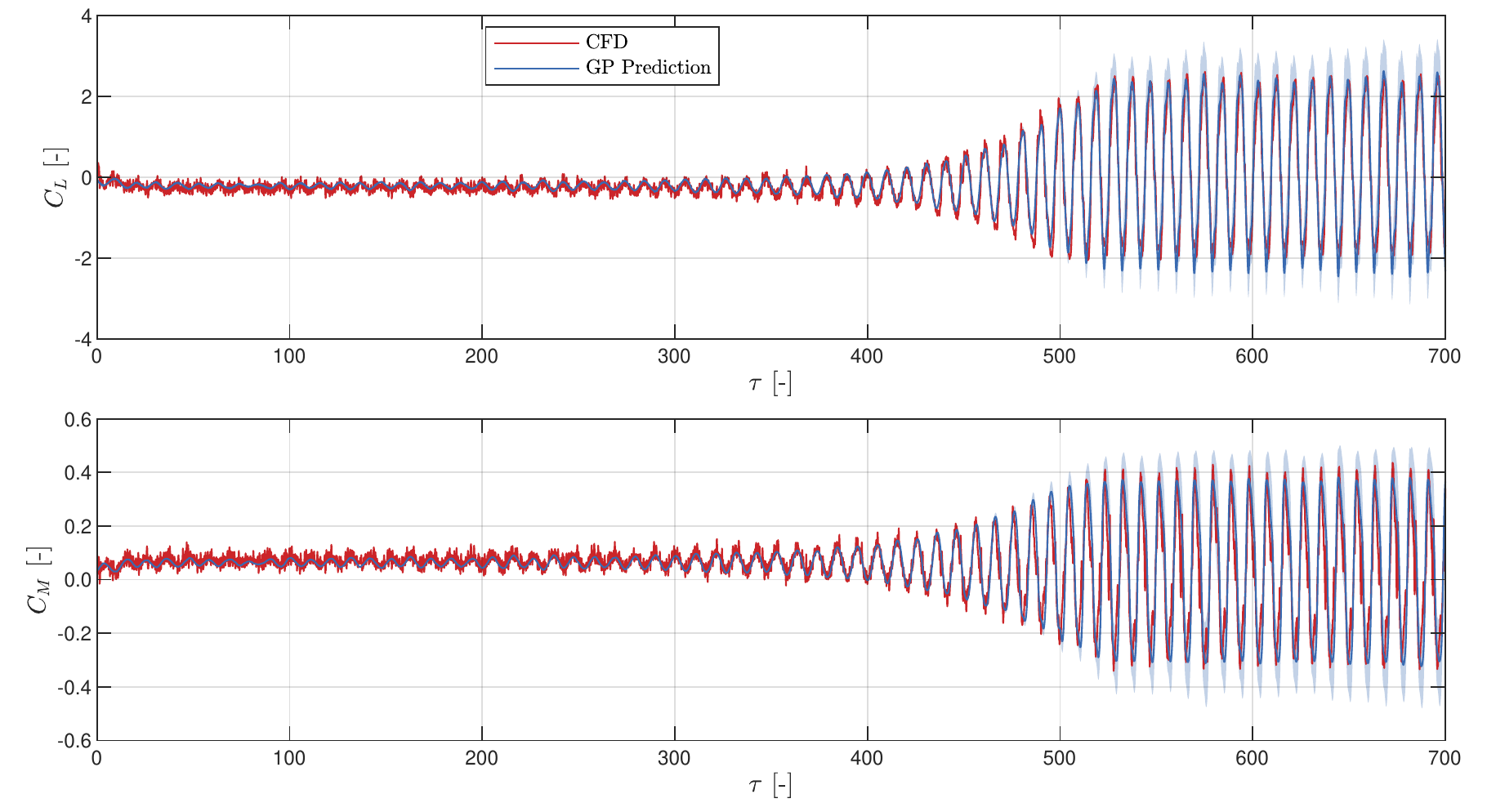} 
	\caption{Great Belt Bridge: Prediction of the lift $C_L$ (top) and moment $C_M$ (bottom) wind coefficients for forced vibration with the CFD displacements from the flutter analysis at $U$=72 m/s as input (cf. Fig.~\ref{fig:Ex2b_Flut}, top). The mean of the GP posterior is represented by the solid line, while the shaded area represents the 99\% confidence interval.}
	\label{fig:Ex2a_ForcedFlutter}
\end{figure*}

\begin{figure*}[!t]
	\centering
	\includegraphics[clip]{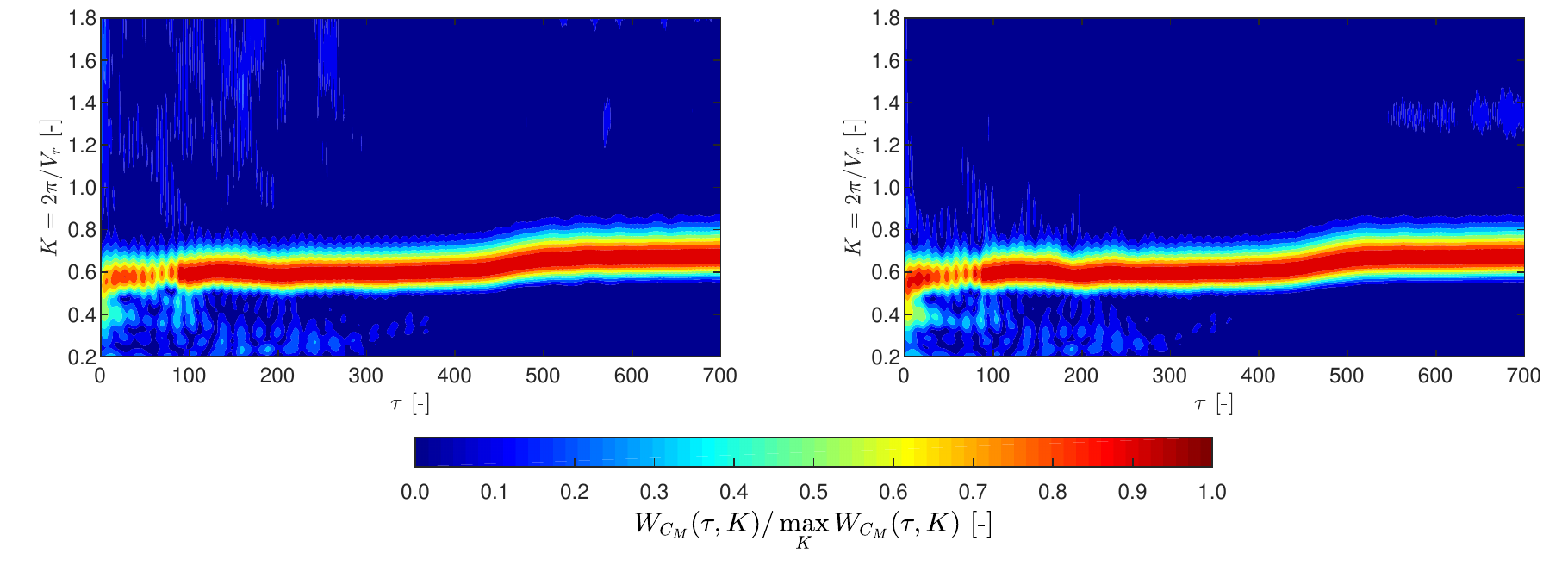} 
	\caption{Great Belt Bridge: Frequency-normalized wavelet magnitude of the moment coefficient $C_M$ from forced vibration based on CFD flutter displacements as an input (cf. Fig.~\ref{fig:Ex2a_ForcedFlutter}, bottom): CFD model (left), and GP model (right).}
	\label{fig:Ex2a_WaveletFlutForce}
\end{figure*}

\begin{figure}[!t]
	\centering
	\includegraphics[clip]{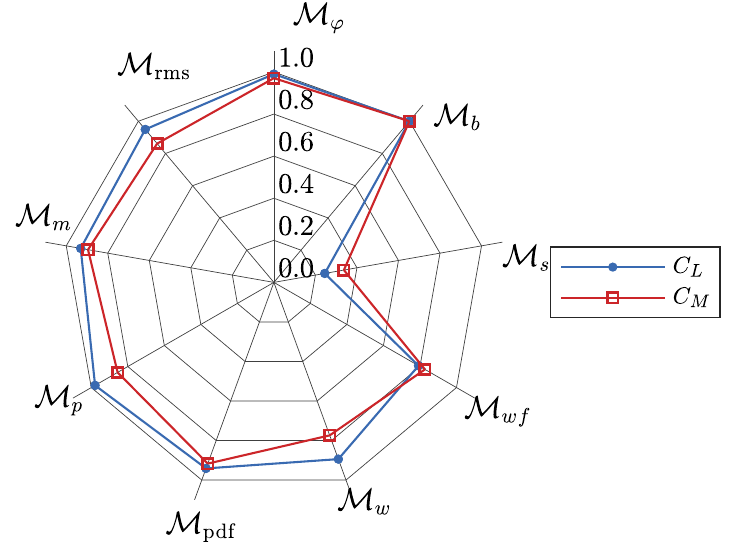} 
	\caption{Great Belt Bridge: Comparison metrics $\mathcal{M}^{\mathrm{CFD},\mathrm{GP}}$ for the lift $C_L$ and moment $C_M$ coefficients for forced vibration based on the CFD flutter displacements as input (cf. Fig.~\ref{fig:Ex2a_ForcedFlutter}). The wavelet central reduced frequency is $K_c$=3$\pi$ and the phase normalization time is $T_c$=5.}
	\label{fig:Ex2a_MetricForcedFlut}
\end{figure}

\subsubsection{Prediction: Hard flutter}
Having verified the GP model's prediction in the previous section, we move on to perform dynamic analysis to predict the critical flutter limit and examine the post-flutter behavior. The dynamic properties of the 2 DOF system are given in Tab.~\ref{tab:NumParam}. Again, only the mean of the GP predictive posterior is used for the dynamic simulation in a one-step-ahead manner, ignoring the uncertainty propagation in the dynamic system. \par
Figure~\ref{fig:Ex2b_Flut} depicts the displacements of the deck undergoing free oscillations for several wind speeds. It can be seen from the figure that both CFD and GP models exhibit coupled flutter for a wind speed of $U_{cr}$=72 m/s, which is similar to the one reported by~\cite{Larsen} from wind tunnel experiments on a section model. The oscillations below this wind speed are damped for the GP model, while the CFD model exhibits small oscillations due to the interior noise. Both models have similar mean displacements for wind speeds below the critical one. At the critical wind speed, the CFD model resulted in a so-called "hard flutter". After the deck is displaced due to the mean wind, coupled flutter initiates and the displacements grow violently. Due to separation at the leading edge, the flutter changes towards the torsional behavior finally reaching LCOs (cf.~\cite{KavrakovMorgenthal2} for further discussion). In the case of the GP model, such behavior is not observed. Instead, the GP self-excited force model yields negative aerodynamic damping at $U_{cr}$ that cancels out the positive structural damping. This results in critical oscillations at $U_{cr}$ with zero effective damping. However, at slightly higher wind speed ($U$=75 m/s), LCO is observed for the GP model. This shows that the self-excited GP forces (i.e. the aerodynamic damping) are nonlinear in an aeroelastic simulation. However, the amplitudes are underestimated by the GP model, leading to the conclusion that nonlinearity is captured to a certain extent.\par
During the LCO for the CFD model, the coupled oscillation frequency is modulated towards the torsional frequency, which can be seen in the frequency-normalized wavelet coefficients of the rotation $W_{\alpha}$ in Fig.~\ref{fig:Ex2b_WaveletFlut} (left). Such transition from coupled towards torsional LCO results from leading-edge separation (cf.~\cite{KavrakovMorgenthal2} for a detailed discussion). The GP model could not capture this phenomenon and results in a constant coupled frequency. However, the figure shows that the initial coupled frequency is similar for the two models ($\tau\approx$0-100 for the CFD model).\par
We performed a multi-step-ahead forced oscillation analysis for the GP model to study this further, taking the oscillations based on the CFD model during flutter as an input. The mean lift and moment time-histories of the GP predictive posterior, including the 99\% confidence interval, are compared to the CFD results in Fig.~\ref{fig:Ex2a_ForcedFlutter}. Generally, the mean GP and CFD forces correspond well, with the CFD being within the GP model's confidence interval and mostly matching the mean. The frequency-normalized wavelet coefficients of the moment $W_{C_M}$ (cf. Fig.~\ref{fig:Ex2a_WaveletFlutForce}) show similar behavior for both models: the mean of the GP posterior exhibits a frequency modulation too. Moreover, Fig.~\ref{fig:Ex2a_MetricForcedFlut} depicts the comparison metrics for the discussed time-histories, taking the CFD as a reference model. Good to fair correspondence can be seen for most of the metrics as most of them resulted in values $\mathcal{M}\geq$0.8, with only the stationary metric $\mathcal{M}_s$ being low. The latter is a consequence of a mismatch in the nonstationary portion of the wavelet coefficients, which is mainly attributed to the interior noise and the minor contribution of the third-order harmonic that appears at $K\approx$1.4 for the GP model (cf. Fig.~\ref{fig:Ex2a_WaveletFlutForce}, right). However, both signals are determined to be nonstationary.\par
Two potential explanations on why the GP model does not capture the nonstationary frequency modulation could be the inappropriate design of experiments for training or model inadequacy. In the former case, the training signal is stationary in the frequency plane, and it may be that the training input space is insufficient to capture the frequency modulation. In the latter case, the model is a finite impulse response model (i.e. NFIR), incapable of capturing frequency shifts, as discussed by~\cite{Schoukens}. Further investigations are warranted on this account, including wind tunnel tests that show such frequency modulations, since the comparison is based on the premise that the CFD simulations are reliable.

\subsection{H-shaped Bluff Deck}
\begin{figure*}[!t]
	\centering
	\includegraphics[clip]{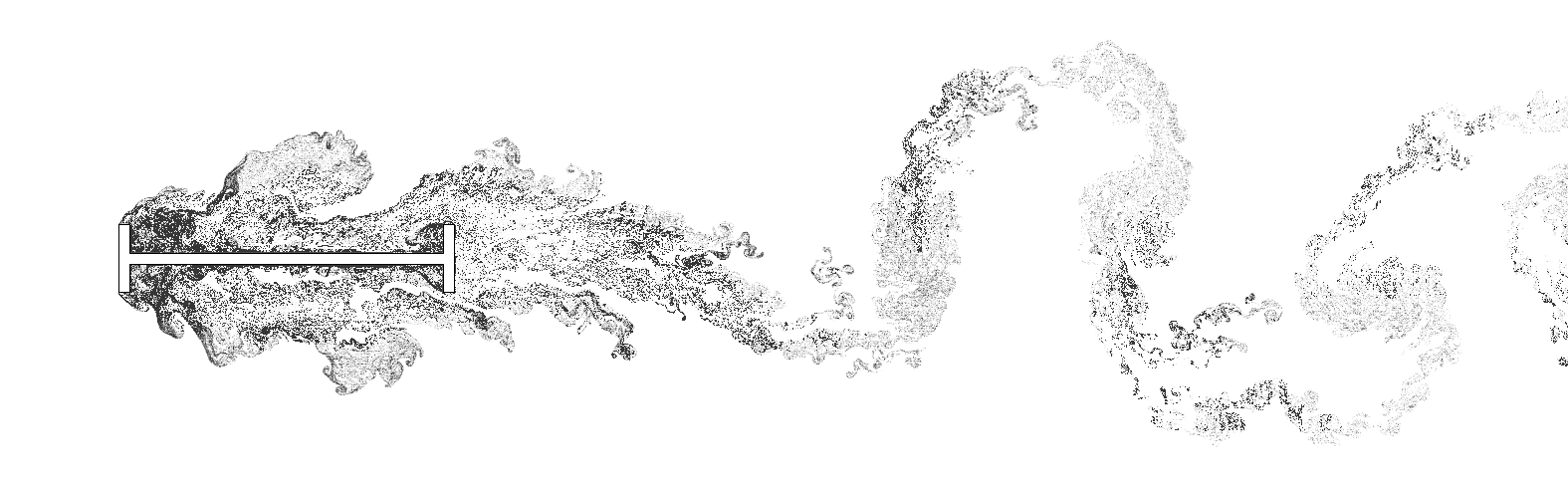} 
	\caption{H-shaped Bluff Deck: Instantaneous particle map from a CFD simulation.}
	\label{fig:Figure_Hsection_Particles1}
\end{figure*}

We apply the presented framework to a simplified H-shaped bridge deck (cf. Fig.~\ref{fig:Schematic_TCGB} and~\cite{AbbasANN} for deck details) to study its performance for bluff bridge decks. The section resembles the Tacoma's deck; however, the aerodynamic behavior of the present section is different than Tacoma's since the longitudinal floor beams, horizontal flanges of vertical beams, shape of the roadway and sidewalks are omitted~\citep{Larsen1998Tacoma}. The parameters for the CFD simulations, training and construction of the GP-NFIR model are given in Tab.~\ref{tab:NumParam}. The section in the CFD simulations is discretized on 330 panels and Fig.~\ref{fig:Figure_Hsection_Particles1} depicts an instantaneous particle map as an example. The system is modeled using 2DOF, despite the common knowledge that H-shaped decks typically experience torsional flutter.

\subsubsection{Learning}
\begin{figure*}[!t]
	\centering
	\includegraphics[clip]{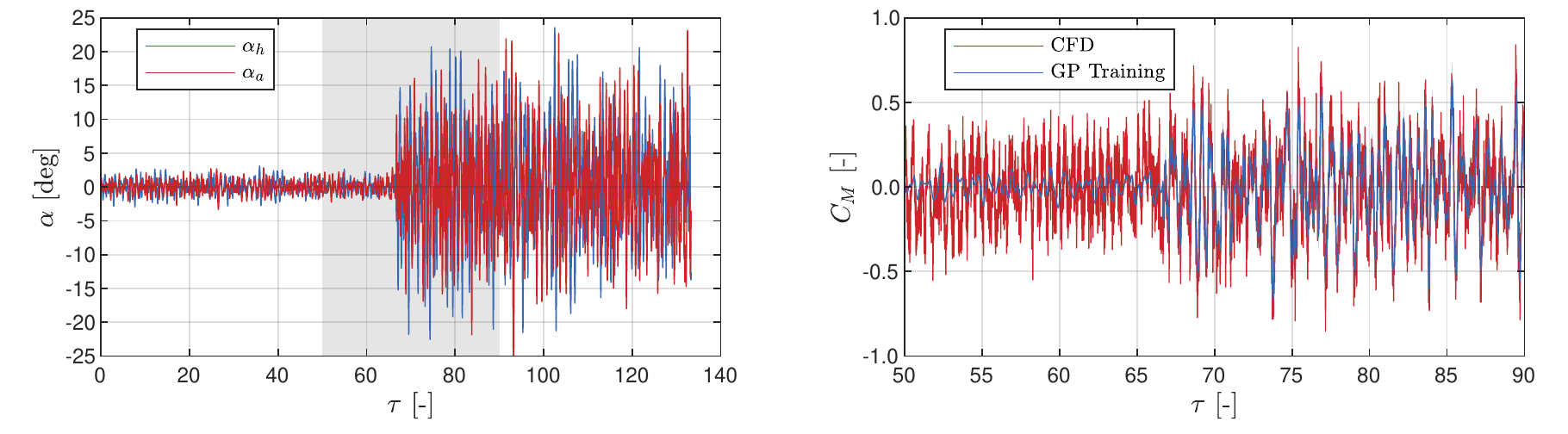} 
	\caption{H-shaped Bluff Deck: Training. Input random excitation for both degrees of freedom $\alpha_h$ and $\alpha_a$ (left) and output moment coefficients $C_M$ (right). The moment coefficients is depicted only for the corresponding shaded area of the input. The mean of the GP posterior is depicted with a solid line, while the 99\% confidence interval is small and thus, not visible. The training is two amplitudes with $\sigma_{\alpha_a}=\sigma_{\alpha_h}$=1 deg and $\sigma_{\alpha_a}=\sigma_{\alpha_h}$=8 deg. The reduced velocity range is $V_r$=1-8.}
	\label{fig:Ex3a_Train}
\end{figure*}

Figure~\ref{fig:Ex3a_Train} depicts the training input motion (left) and a sample of the resulting moment $C_M$ (right) that corresponds to the shaded rectangle in the input. In the GP model case, both the mean and confidence interval of the posterior are given; however, due to the low uncertainty, the confidence interval is hardly visible. Unlike for the Great Belt's deck, here, the interior noise is so violent that the CFD moment coefficient amplitude is roughly the same for both ranges of amplitude ($\sigma_{\alpha1}$=1 deg and $\sigma_{\alpha2}$=8 deg). The violent vortex shedding is due to deck's bluffness and massive separation at the leading edges (cf. Fig.~\ref{fig:Figure_Hsection_Particles1}). As the only input for the GP model is the motion, the interior-noise forces cannot be captured; thus, the GP mean prediction sort of filters the self-excited forces from $C_M$ for the CFD model. How well are the self-excited forces replicated is discussed subsequently. 

\subsubsection{Prediction: Forced vibrations}
The GP model's predictive quality is assessed here for small amplitude forced oscillation (i.e. linear aerodynamics). We only look at the moment coefficient $C_M$ for sinusoidal rotation and, correspondingly, the flutter derivative related to the torsional damping $A_2^*$. This derivative plays a crucial role in a torsional flutter. The sinusoidal forcing amplitude and reduced velocity range are given in Tab.~\ref{tab:NumParam}. Again, the GP predictive distribution is obtained from a multi-step-ahead analysis.\par
\begin{figure*}[!t]
	\centering
	\includegraphics[clip]{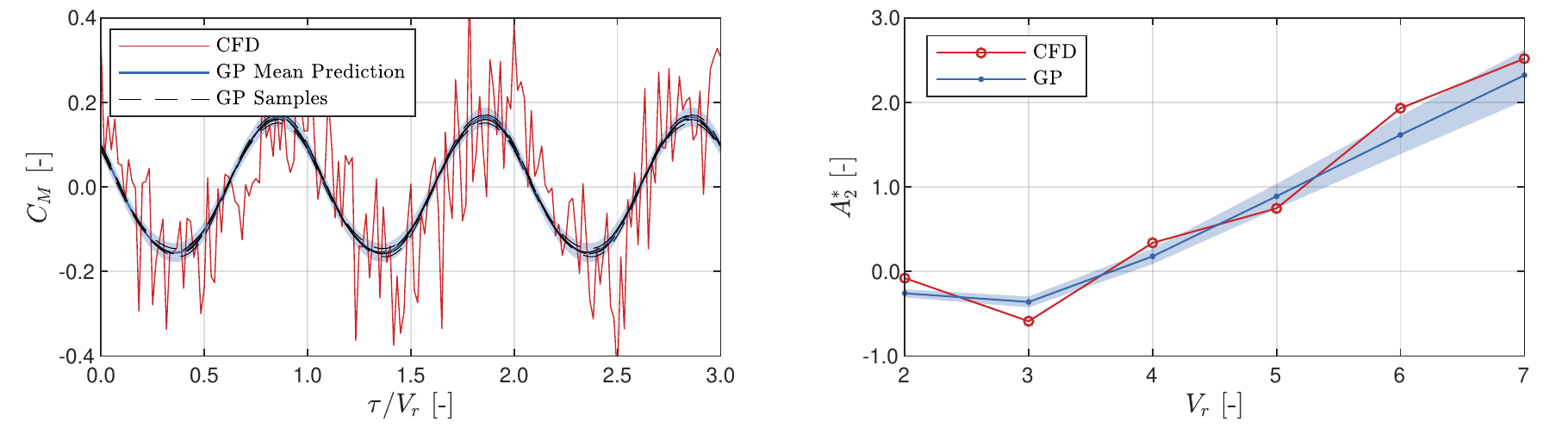} 
	\caption{H-shaped Bluff Deck: Prediction of the moment coefficient $C_M$ for a sinusoidal rotation (amplitude $\alpha_0$=3 deg, $V_r$=4, left), with mean (solid line) and confidence interval (shaded area) for the GP posterior. Prediction of the $A_2^*$ flutter derivative around zero static angle of attack (right), where the mean (solid line) and 99\% confidence interval (shaded area) for the GP model are obtained from 1000 samples that are drawn from the posterior at each $V_r$.}
	\label{fig:Ex3a_FlutterDer}
\end{figure*}

Figure~\ref{fig:Ex3a_FlutterDer} (left) depicts the resulting moment coefficient $C_M$ for an input motion with reduced velocity of $V_r$=4 for both models. For the GP model, the mean and the confidence interval of the predictive distribution are shown, including few random samples. Based on 1000 such random samples, the mean and 99\% confidence interval of the $A_2^*$ flutter derivative around zero static angle of attack are shown in Fig.~\ref{fig:Ex3a_FlutterDer} (right). The $A_2^*$ derivative between the models corresponds good, and the GP model accurately captures the transition from negative to positive damping. Moreover, the flutter derivative's uncertainty interval is significantly larger than one for Great Belt's deck (cf. Fig.~\ref{fig:Ex2a_FlutterDer}) for the GP model. This is mainly due to the high interior noise forces in the training set.\par
\begin{figure}[!t]
	\centering
	\includegraphics[clip]{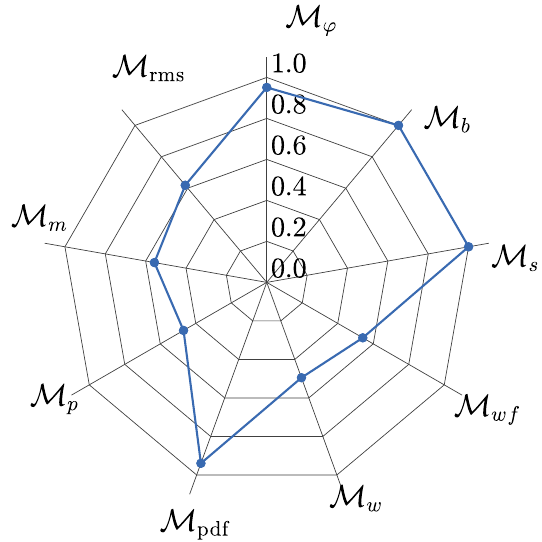} 
	\caption{H-shaped Bluff Deck: Comparison metrics $\mathcal{M}_{C_M}^{\mathrm{CFD},\mathrm{GP}}$ for the moment coefficients $C_M$ from sinusoidal forced rotation for the CFD and mean GP models (cf. Fig.~\ref{fig:Ex3a_FlutterDer}, left). The wavelet central reduced frequency is $K_c$=4$\pi$ and the phase normalization time is $T_c$=2.}
	\label{fig:Ex3a_MetricSine}
\end{figure}
\begin{figure*}[!t]
	\centering
	\includegraphics[clip]{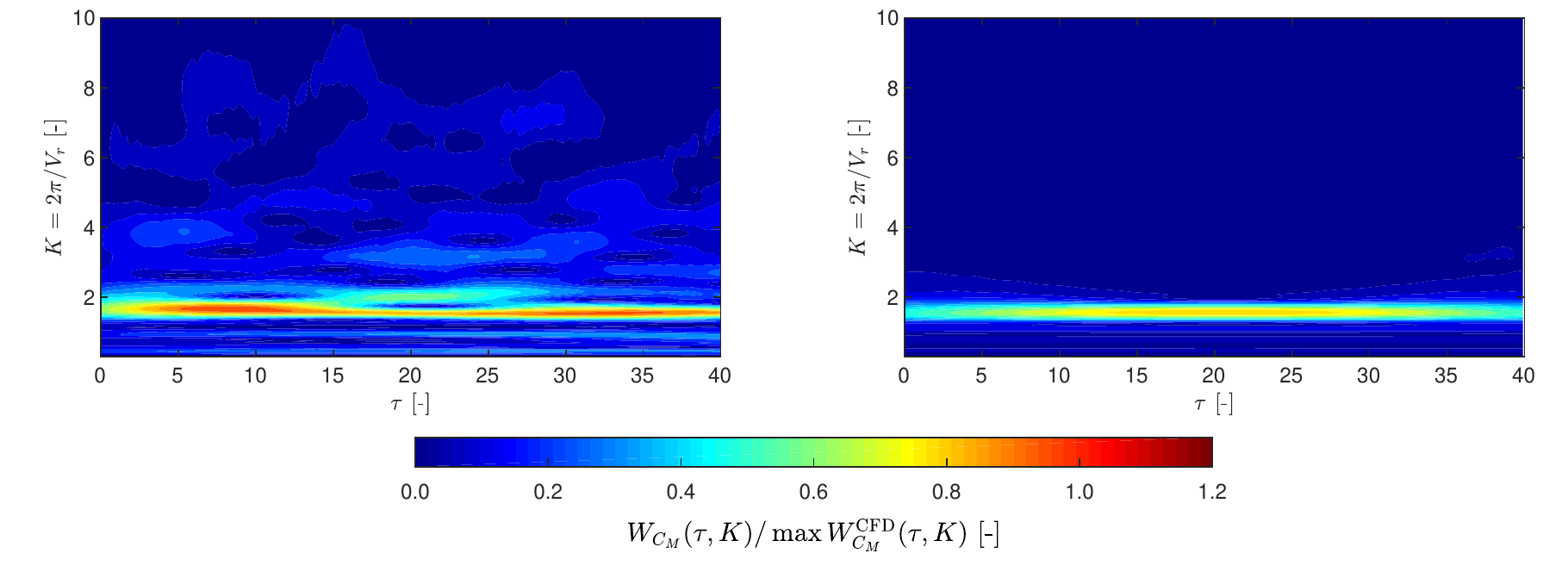} 
	\caption{H-shaped Bluff Deck: Normalized wavelet magnitude of the moment $C_M$ from forced sinusoidal vibration (cf. Fig.~\ref{fig:Ex3a_FlutterDer}, left) for the CFD (left) and GP model (right).}
	\label{fig:Ex3a_WaveletSineForce}
\end{figure*}

Assuming linearity in the motion-force relation is in place, which is generally a valid assumption for such small oscillation amplitudes, the GP model's prediction captures only the main harmonic of the forcing and disregards the violent interior noise (cf.~Fig.\ref{fig:Ex3a_FlutterDer}, right) that causes local nonlinearities. The GP model is incapable of capturing interior noise effects and the associated local nonlinearities. To further analyze these signals, the comparison metrics are given in Fig.~\ref{fig:Ex3a_MetricSine}. Most of the magnitude-related metrics ($\mathcal{M}_\mathrm{rms}$, $\mathcal{M}_\mathrm{m}$, $\mathcal{M}_\mathrm{p}$, $\mathcal{M}_\mathrm{w}$, and $\mathcal{M}_\mathrm{wf}$) resulted in relatively low values due to the additional high-frequency component. This component is also visible in the wavelet coefficients $W_{C_M}$ in Fig.~\ref{fig:Ex3a_WaveletSineForce}. The wavelet coefficients for the CFD model (cf. Fig.~\ref{fig:Ex3a_WaveletSineForce}, left) contain one main harmonic frequency, similar to the forcing frequency, and scattered high-frequency components. The GP model captures only this main harmonic component, and it corresponds well with the CFD model.  \par


\subsubsection{Prediction: Soft flutter}
Having the linear range corresponding well between models, we advance to obtain the critical flutter velocity from a free vibration dynamic analysis and investigate the nonlinear aerodynamics through the post-flutter behavior. The system is modeled as a 2 DOF system and the structural parameters are given in Tab.~\ref{tab:NumParam}.
Again, the mean of the predictive GP posterior is only included in the dynamic analysis, neglecting the uncertainty propagation.\par
\begin{figure*}[!t]
	\centering
	\includegraphics[clip]{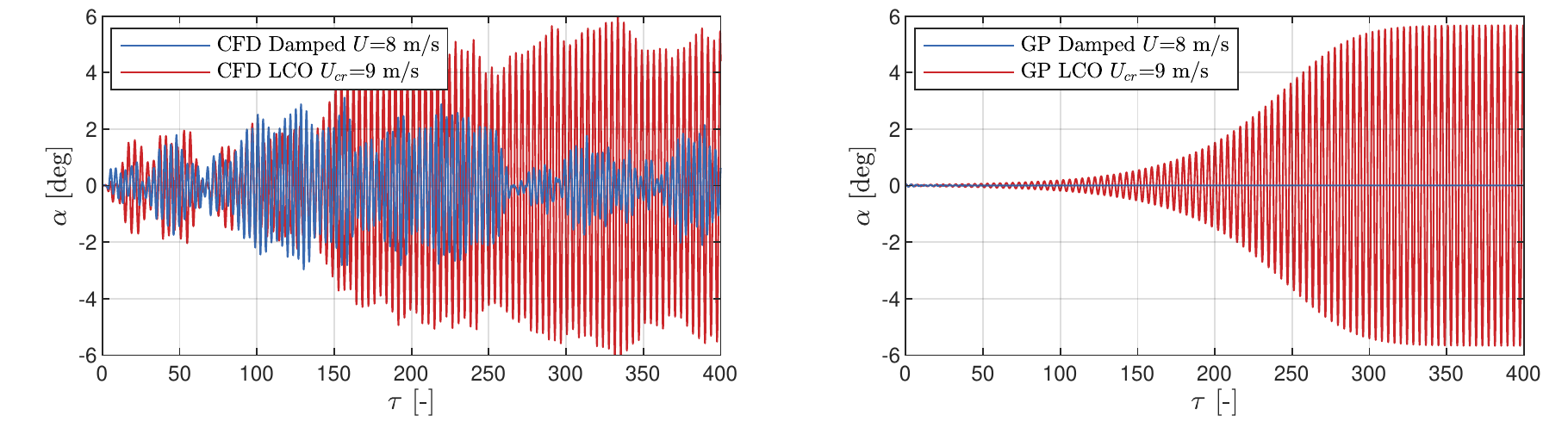} 
	\caption{H-shaped Bluff Deck: LCO prediction of the rotation $\alpha$ from a vree vibration analyses with prescribed initial displacements for the CFD model (left) and the GP model (right).}
	\label{fig:Ex3b_Flut}
\end{figure*}

Figure~\ref{fig:Ex3b_Flut} depicts the rotational displacements for both models for damped oscillations and LCO. The flutter velocity is identical for both models $U_{cr}$=9 m/s. This flutter velocity corresponds to a reduced velocity of $V_{r,cr}\approx$3.75, which is where $A_2^*$ approximately crosses the abscissa (cf. Fig.~\ref{fig:Ex3a_FlutterDer}, left). The flutter limit also compares good with other studies ($U_{cr}$= 7.8 m/s, cf.~\cite{Larsen1998Tacoma}), noting that the structural damping may be different. The corresponding flutter velocities show that the GP model can accurately predict a single DOF flutter. Moreover, the post-flutter behavior is similar for both models, particularly with corresponding LCO amplitude (approx. 5.5 deg). Compared to Great Belt's, the H-shaped deck experiences a so-called "soft flutter" in the CFD analyses. Although there is no clearly defined distinction between soft and hard flutter, in the case of a soft flutter, the increment in the displacement amplitude is relatively small, but distinct. Further experimental validation on this type of behavior for H-shaped sections is warranted since studies of the original Tacoma Narrows deck~\citep{Matsumoto2003} reported mixed vortex-induced vibration and torsional flutter; however, this is beyond the purpose of this study. With this, it can be concluded that the presented GP model accurately captures the nonlinear aerodynamic behavior during LCO for a soft flutter.

\section{Conclusion}
A GP-based data-driven model of the self-excited aerodynamic forces was presented in this paper. The model was formulated as an NFIR model with the effective angle of attack and its memory as exogenous input, taking an entirely nondimensional form from an aerodynamical perspective. The latent function relating the input to output was assumed to be a GP, and the training input was designed as a band-limited random harmonic signal. The fundamental application to a flat plate and the practical application to an H-shapedd and Great Belt's simplified decks showed that the presented model could replicate both linear and nonlinear aerodynamics and successfully predict coupled torsional flutter. Moreover, the post-flutter behavior during LCO was captured well for the soft flutter and, to a certain extent, for the hard flutter.  \par 
With the current nondimensional form of the presented model, the scaling of the aerodynamic forces, e.g., between experimental and prototype models, is inherent. This form makes the model attractive from a practical perspective. Contrary to most semi-analytical aerodynamic models (e.g., quasi-steady or linear unsteady), the presented data-driven model has the mathematical capabilities to capture a complex aerodynamic behavior such as higher-order harmonics, amplitude-dependence, and nonlinear fading fluid memory. Compared to the CFD models, the presented model is a reduced-order model and, once trained, it requires substantially less computational time for a prediction. \par 
Setting up the model is relatively straightforward due to the use of GPs as a nonparametric machine learning method for the latent input-to-output function. It requires only selecting the type of covariance function, which hyperparameters are learned, and the number of lag terms. Selecting up a model structure, such as the number of layers and neurons for neural networks, is avoided. Moreover, overfitting is avoided as GPs abide Occam's razor during training. However, it did not capture the frequency modulation during hard flutter for the Great Belt's deck and the interior noise, including the associated local nonlinearities. This could possibly be avoided by a more careful design of experiments for training and model formulation, which are still open questions in data-driven aerodynamics.  \par
In conclusion, the presented GP-NFIR model shows the potential for both more profound insight into nonlinear aerodynamics and practical applications, such as the structural design of long-span bridges and tall towers. The presented procedure allows the training data to be based on experiments, CFD analyses, or both combined in a hybrid manner to reduce experimental and numerical uncertainty. \par 
The framework is by no means complete. Further analyses are required to determine the number of lag terms or appropriate GP kernels that fulfill physical relations in structural aerodynamics. Additionally, validation using data from wind tunnel testing would enhance the validity of the presented model and show its applicability for noisy experimental measurements. The model could be extended to include the drag and lateral motion for 3D analyses in a straightforward manner. Expanding the framework online learning, based on real-time monitoring data, remains a viable outlook as well. This extension would potentially increase this framework's utility for structural control, maintenance, and life-cycle analysis. \par
\section*{Data Statement}
The Matlab code, including the results from Sec.~\ref{sec:FundApp}, is publicly available online: \href{https://github.com/IgorKavrakov/AeroGP}{github.com/IgorKavrakov/AeroGP}.
\section*{Author contributions}
Igor Kavrakov: Conceptualization, Methodology, Software, Writing - Original Draft.
Allan McRobie: Conceptualization, Methodology, Supervision, Writing - Review \& Editing. Guido Morgenthal: Conceptualization, Software, Supervision, Funding acquisition, Writing - Review \& Editing. 	

\section*{Acknowledgments}
IK and GM gratefully acknowledge the support by the German Research Foundation (DFG) [Project No. 329120866]. IK gratefully acknowledges the Visiting Fellowship by the University of Cambridge during his stay at the Department of Engineering.

\bibliographystyle{elsarticle-harv}
\bibliography{references}

\end{document}